\documentclass[twocolumn]{aastex63}

\usepackage{graphicx}

\shorttitle{Breaking resonant chains}
\shortauthors{Matsumoto \& Ogihara}

\begin{document}

\title{Breaking Resonant Chains: Destabilization of Resonant Planets due to Long-term Mass Evolution}

\correspondingauthor{Yuji Matsumoto}
\email{ymatsumoto@asiaa.sinica.edu.tw}

\author[0000-0002-2383-1216]{Yuji Matsumoto}
\affiliation{Institute of Astronomy and Astrophysics, Academia Sinica, Taipei 10617, Taiwan}
\author[0000-0002-8300-7990]{Masahiro Ogihara}
\affiliation{National Astronomical Observatory of Japan, 2-21-1, Osawa, Mitaka, 181-8588 Tokyo, Japan}

\begin{abstract}
	Recent exoplanet observations reported a large number of multiple-planet systems, in which some of the planets are in a chain of resonances. 
	The fraction of resonant systems to non-resonant systems provides clues about their formation history. 
	We investigated the orbital stability of planets in resonant chains by considering the long-term evolution of planetary mass and stellar mass and using orbital calculations. 
	We found that while resonant chains were stable, they can be destabilized by a change of $\sim$10\% in planetary mass. 
	Such a mass evolution can occur by atmospheric escape due to photoevaporation. 
	We also found that resonant chains can be broken by a stellar mass loss of $\lesssim1$\%, which would be explained by stellar winds or coronal mass ejections. 
	The long-term mass change of planets and stars plays an important role in the orbital evolutions of planetary systems including super-Earths.
\end{abstract}


\section{Introduction}\label{sect:intro}

Recent observations have revealed that the orbital architecture of exoplanet systems are composed of multiple planets in close-in orbits \citep[e.g.,][]{Fabrycky+2014,Weiss+2018}.
Almost all the members of these systems have smaller radii than 4 Earth radii, thereby suggesting that these are not giant planets but sub-Neptunes or super-Earths.
Their orbital periods are within $\sim 100$ days \citep{Weiss+2018b}.
Their orbital distribution provides insights about their formation history.
In particular, the information about orbital resonances provides useful constraints \citep[][]{Ogihara+2018}. 
Although several planetary systems are in chains of resonances \citep{Mills+2016, MacDonald+2016,Gillon+2017}, most multiple-planet systems are not \citep{Fabrycky+2014,Winn&Fabrycky2015}.
The number of planets in resonant chains are between four and seven, which is larger than the average number of Kepler planets \citep[$3.0\pm0.3$,][]{Zhu_W+2018}.

Theoretical studies showed that planets are trapped in resonant chains through orbital migration \cite[e.g.,][]{Terquem&Papaloizou2007,Ogihara&Ida2009}.
The number of planets in resonant chains determines whether such resonant chains will remain for long periods of time.
When the number of planets is large, the orbit crossing time is short, thereby leading to breaking the chain after gas dispersal. 
\cite{Matsumoto+2012} showed that there exists a critical number for orbital instability. 
When the number of planets in resonances is smaller than the critical number ($\sim 10$), which depends on orbital properties, the system can be significantly stabilized.
Compared with non-resonant systems, the orbit crossing time becomes longer by several orders of magnitude. 

Several studies have focused on reproducing the small fraction of resonant chains in observed super-Earth and sub-Neptune systems. 
\cite{Izidoro+2017} performed 120 $N$-body simulations and showed that the fraction of systems in resonant chains is larger than that of super-Earths observed by {\it Kepler}. 
In addition, the typical number of planets in resonant chains is not consistent with observations. 
In their simulation results, the number of planets in resonant chains is typically between 6 and 10, which is larger than the typical number of planets in observed super-Earths (i.e., between four and seven). 
These discrepancies indicate that resonant chains can be destabilized even when the number of planets in the chain is smaller than the critical number obtained in the previous study of \cite{Matsumoto+2012}. 
Some additional mechanisms likely play roles in breaking resonant chains.

In this study, we investigate the effect of long-term mass evolution of planets and stars.
Close-in planets that grow in the gas disk would accrete H/He atmospheres that come from the protoplanetary disk \citep{Ikoma&Hori2012}.
These atmospheres escape from planets by the photoevaporation \citep[e.g.,][]{Valencia+2010, Lopez&Fortney2013,Owen2019,Hori&Ogihara2020}, core-powered mass loss \citep[][]{Ginzburg+2016, Gupta&Schlichting2019}, and Parker wind \citep[][]{Owen&Wu2016}.
Planets can lose $\gtrsim$10\% of their masses through the atmospheric escape.

The stars also lose their masses.
Currently, the Sun loses its mass via the solar wind \citep[e.g.,][]{McComas+2000} and coronal mass ejections \citep[e.g.,][]{Munro+1979, Jackson&Howard1993, Yashiro+2006}.
Observations showed that the stellar mass loss rate increases as the stellar magnetic activity increases \citep{Wood+2002,Wood+2005,Gundel2004}.
Stars have large mass-loss rates in their young ages since their magnetic activities are stronger \citep{Ribas+2005, Aarnio+2012, Suzuki+2013}.
Stars can lose $\sim$~0.1\%--1\% of their mass in the first 1~Gyr \citep{Wood+2002}.
The stellar mass loss is known as a possible solution to the faint young Sun paradox \citep{Sagan&Mullen1972, Feulner2012}.

Such mass evolutions affect the orbital stability of planets in resonant chains. 
Previous studies of the orbital stability of planets not present in resonant chains showed that the orbital crossing timescale was a decreasing function of the mass ratio between the planets and central star \citep{Chambers+1996,Zhou+2007}. 
Similar dependencies are obtained for resonant chains \citep[][]{Matsumoto+2012}. 
This suggests that resonant chains can be destabilized when stars lose their masses. 
Moreover, the orbital stability of planets in resonant chains would be affected by the evolutions of semimajor axes induced by the stellar mass evolution \citep{Minton&Malhotra2007}.
However, it is not clear whether such small changes in masses ($\sim$0.1\% -- 10\%) will make the planetary system in resonant chains unstable. 
Instead, the planetary mass loss may further stabilize resonant chains because the orbital separation scaled by the Hill radius becomes larger.

We have investigated the orbital instability of planets in resonant chains by considering the long-term mass evolution of planets and stars.
We consider the mass evolution of planets and stars separately. 
We show that their mass loss events make the resonant planets unstable especially when the number of planets is close to the critical number \citep{Matsumoto+2012}.
The structure of this paper is as follows.
Our numerical model is described in Section \ref{sect:model}.
We show our results for the planetary mass evolution in Section \ref{sect:result_pl} and those for stellar mass evolution in Section \ref{sect:result_st}.
Our conclusions are presented in Section \ref{sect:conclusion}.

\section{Model}\label{sect:model}

\subsection{Overview}

We follow \cite{Matsumoto+2012} to calculate the orbital crossing time of planets in the chain of resonances.
At first, we form the system of planets in the chain of $p+1$:$p$ resonances through orbital migration in a gaseous disk. 
Then, we calculate orbital crossing times of these systems by $N$-body simulations, including gas depletion, using the depletion timescale $t_{\rm dep}$.
The first one is called the capture simulation, and the latter one is the stability simulation.

\begin{deluxetable*}{ccccc}
	\tablenum{1}
	\tablecaption{Models}\label{table:models}
	\tablehead{
		Model	&	initial planetary mass	&	resonance	&	initial orbital separation	&	critical number	\\
				&	($M_{\rm init}/M_{\odot}$)	&	($p+1:p$)	&	($\Delta a/r_{\rm H,init}$)		&	($N_{\rm crit}$)
	}
	\startdata
		$\mu$4$p$3	&	$10^{-4}$	&	4:3	&	4.72	&	6\\
		$\mu$4$p$2	&	$10^{-4}$	&	3:2	&	6.63	&	7\\
		$\mu$4$p$1	&	$10^{-4}$	&	2:1	&	11.2	&	9\\
		$\mu$5$p$5	&	$10^{-5}$	&	6:5	&	6.45	&	7\\
		$\mu$5$p$4	&	$10^{-5}$	&	5:4	&	7.89	&	6\\
		$\mu$5$p$3	&	$10^{-5}$	&	4:3	&	10.2	&	13
	\enddata
	\tablecomments{
		Summary of our models.
		In our models, the initial planetary mass ($M_{\rm init}/M_{\odot}$) and resonant value $p$ are parameters.
		The initial orbital separations ($\Delta a/r_{\rm H,init}$) are derived from these parameters and are shown in this table.
		The critical number of planets in each model ($N_{\rm crit}$) is our numerical results.
		When the number of planets in the resonant chains is equal or less than $N_{\rm crit}$, planets do not cause orbital instability in our simulations without mass evolutions.
	}
\end{deluxetable*}

\begin{figure}[ht]
	\plotone{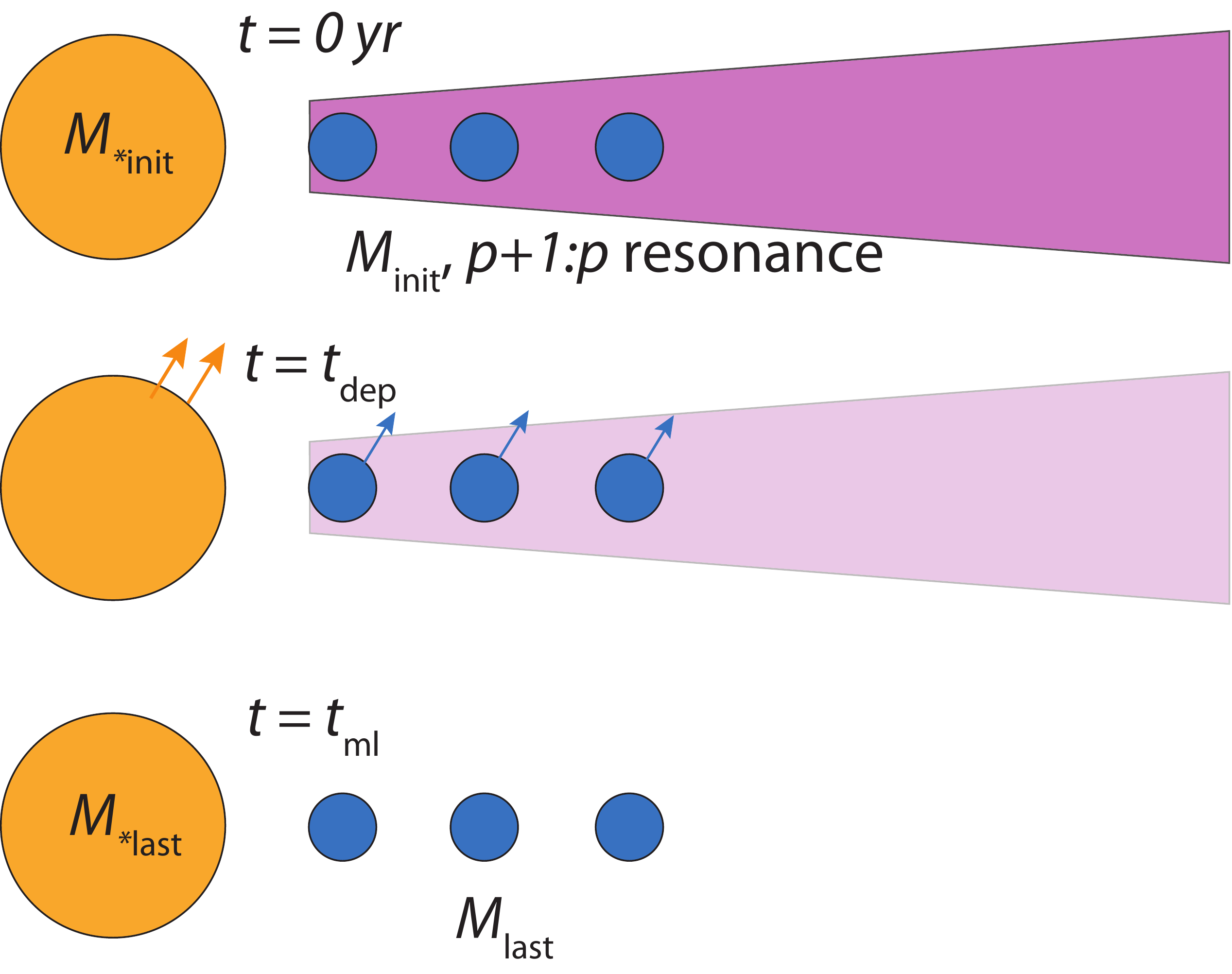}
	\caption{
		The schematic figure of the stability simulations.
		Protoplanets with the mass of $M_{\rm init}$ are in $p+1$:$p$ resonant chains around the star with the mass of $M_{\rm *init}$ in the gas disk.
		The gas disk dissipates in $t_{\rm dep}=10^3$~yr.
		Either planetary mass or stellar mass evolves with $t_{\rm ml}$.
		In the case of planetary mass evolution, the final mass of the planets is $M_{\rm last}$.
		In the stellar case, the final mass of the star is $M_{\rm *last}$.
	}
	\label{Fig:schematic}
\end{figure}

In the stability simulations, we consider two models: one is the planetary mass evolution, and the other is the stellar mass evolution.
The schematic picture of the stability simulations is shown in Figure \ref{Fig:schematic}.
Planets initially have the same mass ($M_{\rm init}$) around the central star whose initial mass is 1 solar mass ($M_{\rm *init}=1M_{\odot}$).
The initial planetary mass is taken as a parameter, $M_{\rm init}=10^{-5}M_{\odot}$ and $10^{-4}M_{\odot}$.
Either planetary mass or stellar mass evolves with $t_{\rm ml}$.
The innermost planet is located at 0.1~au and the semimajor axes of the other planets are given by the $p+1:p$ resonances.
According to our setting for the resonant values $p$ and $M_{\rm init}$, the orbital separations normalized by the Hill radius ($\Delta a/r_{\rm H}$ where $r_{\rm H}=[ 2M/(3M_{*}) ]^{1/3} (a_i+a_{\rm i+1})/2 $, $M$ is the planetary mass, $a_i$ is the semimajor axis of the $i$-th innermost planet, $M_*$ is the stellar mass) become the equal values \citep[e.g.,][]{Matsumoto+2012, Weiss+2018}.
The Hill radius changes according to the mass evolution of planets or stars as $r_{\rm H}=r_{\rm H, init} \mu^{1/3} \mu_{\rm init}^{-1/3} $ where $\mu=M/M_*$ and $\mu_{\rm init}=M_{\rm init}/M_{\rm *init}$.
All planets share their orbital planes.
Our parameters are summarized in Table \ref{table:models}.
Our models are named from the initial planet-star mass ratio and the resonant value $p$.
In this paper, we mainly focus on the $\mu4p3$ models where $10^{-4}M_{\odot}$ mass planets are in 4:3 resonant chains for a clear presentation.
The results of the other models are presented in the Appendices \ref{sect:reslt_pl_ap} and \ref{sect:reslt_st_ap}.

\begin{deluxetable*}{clc}
	\tablenum{2}
	\tablecaption{Key quantities}\label{table:quantities}
	\tablehead{
		quantities	&	explanations	
	}
	\startdata
	$t_{\rm cross}$	&	The orbital crossing time	\\
	$N$				&	The number of planets\\
	$p$				&	Commensurability of planets $p+1$:$p$ \\
	$M_{\rm li}$ 	&	The ratio between the last and initial planetary mass (equal to $M_{\rm last}/M_{\rm init}$) \\
	$M_{*\rm li}$ 	&	The ratio between the last and initial stellar mass (equal to $M_{*\rm last}/M_{*\rm init}$) \\
	$t_{\rm ml}$	&	The timescale for the planetary mass or stellar mass evolution	($10^4$ yr) \\
	$f_{\rm ml}$	&	The number fraction of planets that experience mass evolution in a system	\\
	$t_{\rm cross,Z07}$		&	The orbital crossing timescale of planets not present in resonances (Equation (\ref{eq:t_Z}))\\
	$t_{\rm dep}$	&	The timescale of disk gas depletion	($10^3$ yr) \\
	$t_{\rm drag}$	&	The timescale for stabilizing planets by disk gas (Equation (\ref{eq:t_drag})) 
	\enddata
\end{deluxetable*}

In stability simulations, we perform calculations until the distance between the planets is less than the Hill radius or the system is stable over $10^7$ yr.
This upper time limit corresponds to $10^{8.5} T_{\rm Kep}$ of the planet at 0.1 au around 1 solar mass star where $T_{\rm Kep}$ is the orbital period.
First, we perform $N$-body simulations fixing the planetary mass and stellar mass as a reference for each model.
We repeat these simulations until we find the critical number of planets in resonances for orbital stability ($N_{\rm crit}$) by increasing the total number of planets in the system ($N$).
We perform three simulations with different initial locations of planets. 
Second, we consider the time evolution of the planetary mass and stellar mass.

We name simulations based on the model (i.e., initial mass and resonant commensurability) and additional parameters (i.e., number of planets and final mass). 
For example, in a simulation called $\mu4p3$\_$N6M_{\rm li}0.9$, six planets with an initial mass of $10^{-4} M_{\odot}$ are in 4:3 resonances and their masses decrease to 0.9 times the initial value at the end of the simulation.
In the following section, we explain our model in detail.
Key quantities are summarized in Table \ref{table:quantities}.

We usually perform one simulation for each case. 
We then choose 26 cases in which $N<N_{\rm crit}$ from $\mu4p3$, $\mu5p5$, and $\mu5p4$ models. 
To account for the chaotic nature of the orbital evolution of multiple planet systems, we perform two additional simulations in 21 cases and four additional simulations in five cases. 
While previous studies suggested that the standard deviation of the logarithm orbital crossing time of planets that are not in resonances is 0.2~dex \citep[][]{Rice+2018, Hussain&Tamayo2020}, standard deviations of 85~\% of our cases with additional simulations are less than 0.2~dex. 
Their median and average values are 0.084~dex and 0.13~dex, respectively. 
The reason why the standard deviation of the crossing time of planets in resonances is smaller than that for non-resonant planets is probably because angular relations of planets in resonant chains are similar when planets are in a resonant chain with small libration angles of their resonant angles.

\subsection{N-body method and migration model}\label{sect:n_body_disk}

The orbits of protoplanets are calculated by numerically integrating the equation of motion.
We adopt the fourth-order Hermite integrator \citep[e.g.,][]{Kokubo&Makino2004} with the hierarchical timestep \citep{Makino1991}. 
We consider the specific forces of eccentricity damping due to tides from the gas disk as a drag force ($\textrm{\boldmath $F$}_{\rm damp}$) and type-I migration ($\textrm{\boldmath $F$}_{\rm mig}$), respectively.
These forces are given by 
\begin{eqnarray}
	F_{{\rm damp},r}		&=& \frac{1}{0.78t_e} \left( 2A_r^c [v_{\theta} -r\Omega_{\rm K} ] + A_r^s v_r \right),\\
	F_{{\rm damp},\theta}	&=& \frac{1}{0.78t_e} \left( 2A_{\theta}^c [v_{\theta} -r\Omega_{\rm K} ] + A_{\theta}^s v_r \right),\\
	F_{{\rm mig},\theta}	&=& \frac{r \Omega_{\rm K}}{2 t_a},
\end{eqnarray}
where $t_e$ is the eccentricity damping timescale, $t_a$ is the migration timescale, $v_r$ and $v_{\theta}$ are radial and azumuthal velocity components, $r$ is the orbital radius, and $\Omega_{\rm K}$ is the Keplerian frequency, respectively.
The numerical factors are given by \citep{Tanaka&Ward2004},
\begin{eqnarray}
	A_r^c = 0.057, && A_r^s = 0.176, \\
	A_{\theta}^c = -0.8686,&& A_{\theta}^s = 0.325.
\end{eqnarray}
For the timescales of eccentricity damping and migration, we follow the formalism of \cite{Tanaka&Ward2004} and \cite{Tanaka+2002},
\begin{eqnarray}
	t_e &=& \left(\frac{f_e}{0.78}\right) \mu^{-1} \left( \frac{\Sigma_{\rm g} r^2}{ M_* } \right)^{-1}
		\left( \frac{c_{\rm s}}{ v_{\rm K} } \right)^4 \Omega_{\rm K}^{-1}
		,
	\\
	t_a &=& \left(\frac{f_a}{ 2.7+1.1q }\right) \mu^{-1} \left( \frac{\Sigma_{\rm g} r^2}{ M_* } \right)^{-1}
	\left( \frac{c_{\rm s}}{ v_{\rm K} } \right)^2 \Omega_{\rm K}^{-1}
	,
\end{eqnarray}
where $\Sigma_{\rm g}$ is a surface density of the gas disk, $q$ is the power law index of the surface density ($q={\rm d}\ln{\Sigma_{\rm g}}/{\rm d} \ln{r}=-3/2$), $c_{\rm s}$ is the sound speed, $v_{\rm K}$ is the Kepler velocity, and $f_e$ and $f_a$ are coefficients.
We adopt a power-law disk similar to the minimum-mass solar nebula model \citep[e.g., ][]{Hayashi1981, Ida&Lin2004a},
\begin{eqnarray}
	\Sigma_{\rm g} &=& 2400 f_{\rm g} \left( \frac{r}{ 1\mbox{~au} } \right)^{-3/2} \quad \mbox{g~cm}^{-2},\\
	c_{\rm s} &=& 1.0\times10^5 \left( \frac{r}{ 1\mbox{~au} } \right)^{-1/4} \quad \mbox{cm~s}^{-1}.
\end{eqnarray}
Although the sound speed depends on the luminosity of the central star in the optically thin disk, we neglect this dependence for simplicity.
The surface density vanishes at the inner edge ($r_{\rm edge}$) with a hyperbolic tangent function of width, $\Delta r=10^{-3}$ au.
In capture simulations, the surface density is constant and $f_{\rm g}=1$.
In stability simulations, where planets are in resonances, we decrease the surface density with time ($t$) as 
\begin{eqnarray}
	f_{\rm g} = \exp{\left( -\frac{t}{t_{\rm dep}} \right)},
\end{eqnarray}
where $t_{\rm dep}$ is the timescale of disk gas depletion.
We take $t_{\rm dep}=10^3$~yr. 
Observations suggested that the disk lifetime and its dissipation time are $\sim 10^6$~yr \citep[][e.g.,]{Haisch+2001, Ribas+2014} and $\sim 10^5$~yr \citep[][]{Williams&Cieza2011}, respectively. 
Therefore, our assumption of $t_{\rm dep}=10^3$~yr is shorter than observationally inferred value. 
Note, however, that it has been shown that when the depletion timescale is longer than the libration timescale of resonant chains and thus the gas depletion is adiabatic, the orbital crossing time does not depend sensitively on $t_{\rm dep}$ \citep[][]{Matsumoto+2012}. 
In addition, it is suggested that the gas in the inner disk dissipates earlier \citep[][]{Ribas+2014}. 
Recent theoretical studies suggest that the gas in the inner disk can be quickly removed by magnetically driven disk winds \citep[][]{Suzuki+2010,Suzuki+2016, Bai&Stone2013ApJ_767_30B}.

We put the planets in $p+1$:$p$ resonant chain from the inner edge by the eccentricity trap and slow migration.
The eccentricity trap is the mechanism by which the planet located at the inner edge receives the angular momentum due to the partial planet-disk interaction in an orbit \citep{Ogihara+2010}.
The eccentricity trap occurs when $t_e$ is much shorter than $t_a$, and $\Delta r/r_{\rm edge}$ is small.
The resonant capture condition is given by \cite{Ogihara&Kobayashi2013}.
When the migration timescale is longer than the critical timescale, planets are trapped in a certain resonance.
According to the $t_a$ and $t_e$ conditions, we take $f_a \geq 50$ and $f_e\geq1$.

\subsection{The evolution of the mass of the star and planets}

In the first $t_{\rm ml}$ of stability simulations, either planetary mass or stellar mass evolves with time.
We take the mass loss timescale ($t_{\rm ml}$) and the mass ratio between the initial and final mass ($M_{\rm li}=M_{\rm last}/M_{\rm init}$ for planets or $M_{\rm *li}=M_{\rm *last}/M_{\rm *init}$ for stars) as our parameters.
The number fraction of the planets that experience the mass evolution ($f_{\rm ml}$) is also our parameter.
In our fiducial cases, we put $f_{\rm ml}=1$, which means that all planets experience the mass evolution.
The exponential function gives the mass evolution.
For the change of planetary mass, we assume 
\begin{eqnarray}
	M &=& M_{\rm init} \exp{ \left[ \frac{t}{t_{\rm ml}} \ln{\left( \frac{ M_{\rm last} }{M_{\rm init}} \right)} \right] }
		= M_{\rm init} \left( \frac{ M_{\rm last} }{M_{\rm init}} \right)^{t/t_{\rm ml}} .
		\nonumber\\
\end{eqnarray}
After $t_{\rm ml}$, $M$ no longer grows and is equal to $M_{\rm last}$.
The stellar mass evolves in the same way.
The parameter range of $M_{\rm last}/M_{\rm init}$ is from 0.5 to 1.5 and that of $M_{\rm *last}/M_{\rm *init}$ is from 0.5 to 1.
The timescale of mass loss takes different values for different mass loss mechanisms. 
We adopt $t_{\rm ml}=10^4$~yr unless otherwise stated. 
As there is a wide range of variations in the mass-loss timescale, we investigate the dependence of the orbit crossing time on $t_{\rm ml}$ in Section \ref{sect:t_ml}. 
As a reference, the mass loss timescale induced by the atmospheric expansion after the disk dispersal is about $10^4$--$10^5$~yr \citep[][]{Owen&Wu2016}. 
Other mechanisms induce a mass loss with longer timescales of about $10^7$--$10^9$~yr \citep[][]{Owen&Wu2017,Gupta&Schlichting2019, Hori&Ogihara2020}.

\section{Results for the planetary mass evolution}\label{sect:result_pl}

\subsection{Typical evolution}

\begin{figure}[ht]
	\plotone{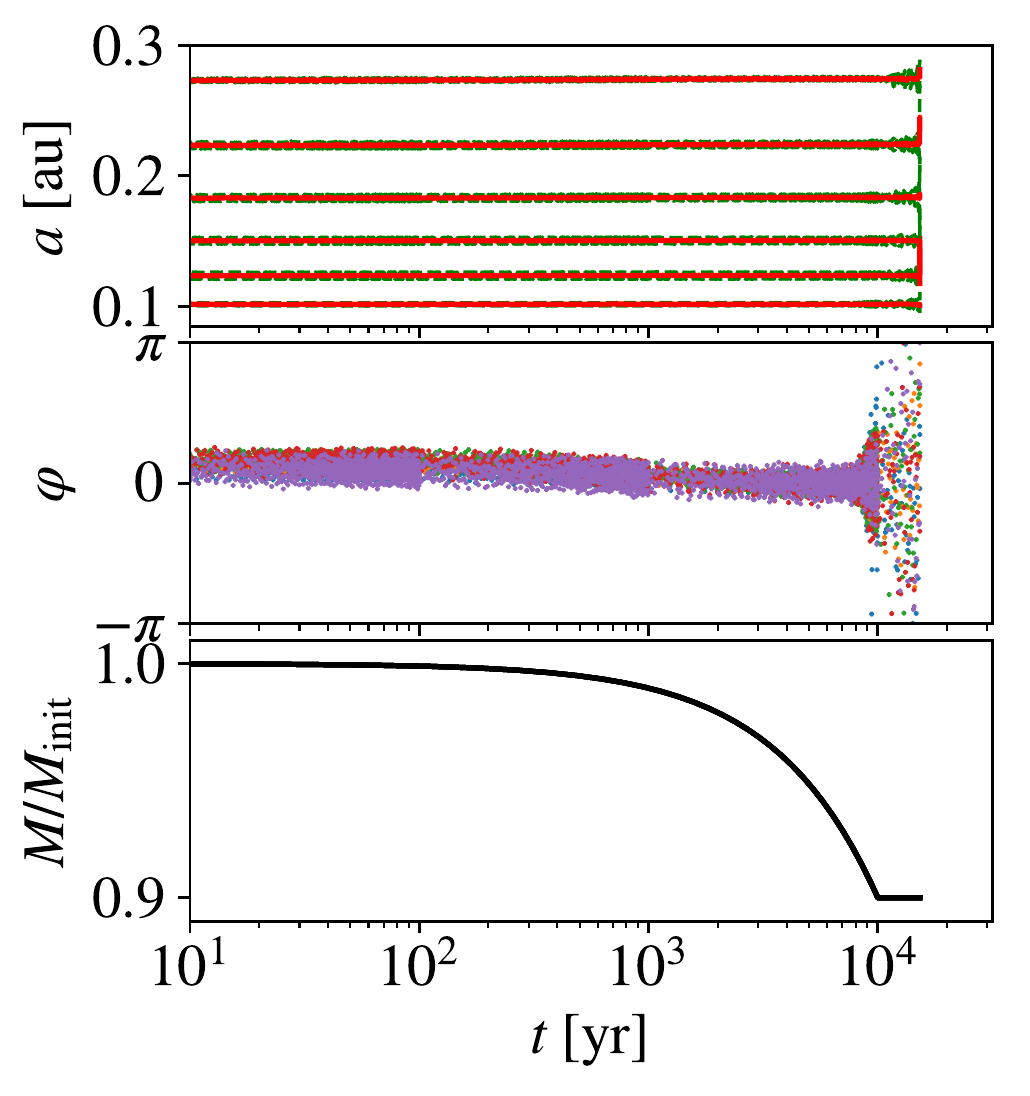}
	\caption{
		The time evolution of planets in the $\mu4p3$\_$N6M_{\rm li}0.9$ case, where six planets whose mass is initially $10^{-4}M_{\odot}$ are in 4:3 resonances.
		These planets lose their mass in the first $10^4$~yr.
		The top panel is the evolution of semimajor axes and pericenter and apocenter distances.
		The semimajor axes are plotted in solid red lines, and pericenter and apocenter distances are in dashed green lines.
		The orbital crossing between the second and third innermost planets occurs at $1.5\times 10^4$ yr.
		The middle panel is the evolution of resonant angles ($\varphi$).
		The bottom panel is the evolution of the planetary mass normalized by the initial value ($M/M_{\rm init}$).
	}
	\label{Fig:t_vs_aMphi_typical}
\end{figure}

At first, we show the typical time evolution of planets with the planetary mass evolution.
Figure \ref{Fig:t_vs_aMphi_typical} shows the time evolution of planets in the $\mu4p3$\_$N6M_{\rm li}0.9$ case. 
Six planets whose mass is initially $10^{-4}M_{\odot}$ are in 4:3 resonances and their final mass is 0.9 times the initial value. 
The masses of planets decrease exponentially, and they become the final values at $10^4$~yr (the bottom panel).
Eccentricities of planets keep their initial values ($\lesssim 10^{-2}$) in the first $\sim 10^4$~yr.
Libration widths of resonant angles begin to increase at $t\simeq 8\times 10^3$~yr and they begin circulations one after another after $10^4$~yr has passed.
Then, eccentricities begin to increase due to secular perturbation.
The second and third innermost planets cause orbital crossing at $1.5\times 10^4$~yr.
It is worth noting that the critical number of planets in this resonant chain is six \citep[][]{Matsumoto+2012}; therefore, these six planets do not cause orbital instability within $10^7$~yr without considering the mass change. 
This indicates that the system is destabilized because of the effect of mass change.
This result is interesting because the orbital separation divided by the Hill radius expands as the planetary mass decreases, which should stabilize the system more \citep[e.g.,][]{Chambers+1996}.

Here, we compare the orbital crossing timescale of $1.5\times 10^4$~yr with other timescales to consider the effect of the resonant chain. 
The orbital crossing is hindered by eccentricity damping due to gas drag \citep[][]{Iwasaki+2001, Iwasaki+2002}.
The orbit stable timescale due to disk gas is $t_{\rm drag} = 8.6\times10^3$~yr in the $\mu4p3$\_$N6M_{\rm li}0.9$ case (see Appendix \ref{sect:stabilization}). 
This explains why the libration width of the resonant angles does not grow in the first $8\times 10^3$~yr. 
The crossing timescale of planets that are not present in resonant chains is $t_{\rm cross,Z07}=1.8\times10^2$~yr (see Appendix \ref{sect:stabilization}). 
This indicates that systems without resonant relationships undergo orbital instability soon after gas depletion. 
The actual orbital crossing time in this simulation is approximately $t_{\rm cross}\simeq 1.7t_{\rm drag}$.
The orbital crossing time is longer than $t_{\rm drag}+t_{\rm cross,Z07}$ due to the resonant effect, i.e., the evolution time of resonant angles from libration to circulation.

\subsection{Dependence on the amplitude of the planetary mass evolution}

\begin{figure*}[hbt]
	\plottwo{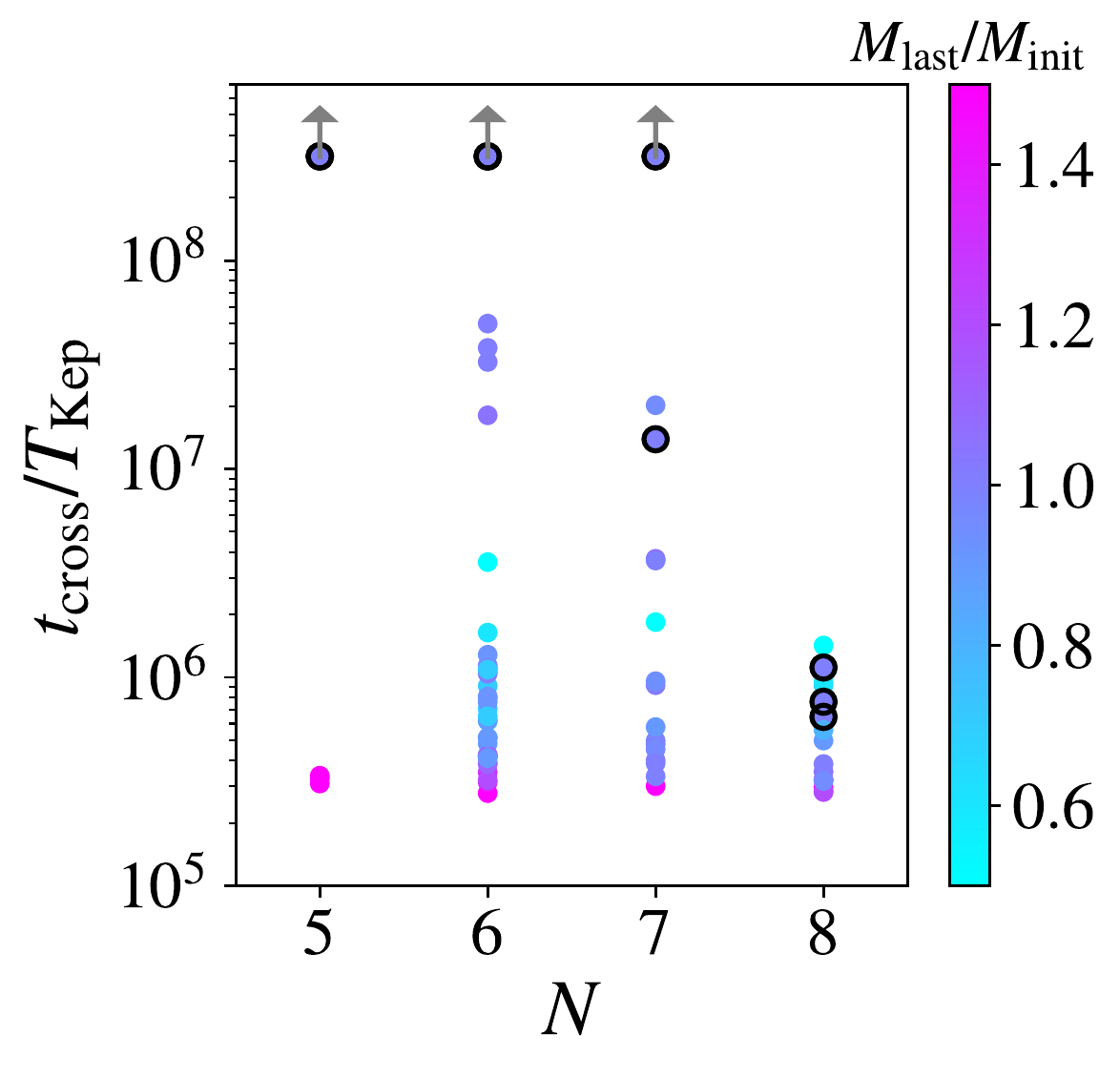}
			{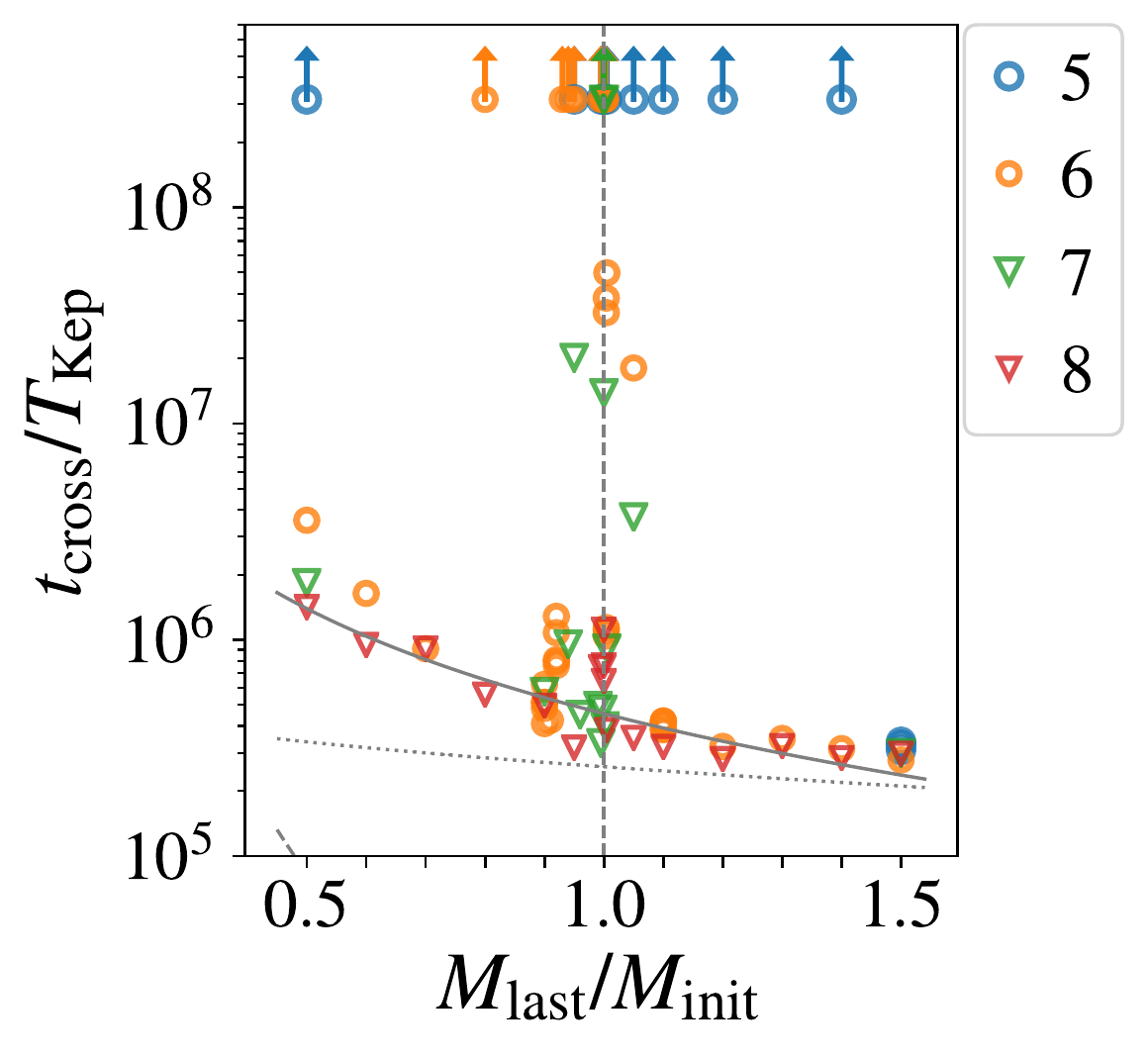}
	\caption{
		The orbital crossing time of planets in the $\mu$4$p$3 model is shown.
		The orbital crossing time is normalized by the Kepler time at 0.1~au ($t_{\rm cross}/T_{\rm Kep}$).
		When the planets do not become unstable in simulations, we plot markers with upper arrows.
		{\it Left}: 
		The orbital crossing time is plotted as a function of $N$.
		The color map is representative of the ratio between the final and initial planet mass ratio ($M_{\rm last}/M_{\rm init}$).
		The points with black edges are the results without mass loss ($M_{\rm last}/M_{\rm init}=1.0$) simulations.
		{\it Right}:
		The orbital crossing time is plotted as the function of $M_{\rm last}/M_{\rm init}$.
		The orbital crossing time of planets in $N\leq N_{\rm crit}$ is plotted as circle markers, and it is triangular in $N> N_{\rm crit}$.
		The vertical dashed line shows $M_{\rm last}/M_{\rm init}=1.0$.
		The dotted line shows $t_{\rm drag}$, and the dashed line shows $t_{\rm cross,Z07}({\tilde e}=0)$.
		The solid line is the fitting line for local short crossing times, 
		$\log{(t_{\rm cross}/T_{\rm Kep})} = -1.6\log{(M_{\rm last}/M_{\rm init})} + 5.6$.
	}
	\label{Fig:num_Mml_t_cross_4_43}
\end{figure*}

We perform simulations changing $N$ and $M_{\rm last}/M_{\rm init}$ to see the dependencies of the orbital crossing time.
The dependence of the orbital crossing time on $N$ in the cases without mass loss is well described by the critical number ($N_{\rm crit}$); the planets in resonant chains are stable in $N\leq N_{\rm crit}$, and they cause orbital instabilities in $N>N_{\rm crit}$ \citep[][e.g., Figure \ref{Fig:num_Mml_t_cross_4_43} in this paper]{Matsumoto+2012}.
The critical number increases as $\Delta a/r_{\rm H,init}$ increases, which is also the case for the mass evolution (Table \ref{table:models}). 
In this section, we explain our results using the $\mu4p3$ model. 
In simulations with different models, we observe similar dependencies on $N$ and $M_{\rm last}/M_{\rm init}$. 
The details of our results in the other models are shown in Appendix \ref{sect:reslt_pl_ap}.

Figure \ref{Fig:num_Mml_t_cross_4_43} shows the orbital crossing time as the functions of $N$ (the left panel) and $M_{\rm last}/M_{\rm init}$ (the right panel) in the $\mu$4$p$3 model.
In the simulations without the mass loss (the points with black edges in the left panel), the planets are always stable for $N\leq 6$. 
That is, $N_{\rm crit} = 6$ in this case.
However, when we consider the planetary mass loss, the planets cause orbital instability even in $N\leq N_{\rm crit}$.
In the $N=5$ case, the planets are stable when $0.5\leq M_{\rm last}/M_{\rm init}\leq 1.4$, and they cause orbital instability when $ M_{\rm last}/M_{\rm init}= 1.5$.
In the $N=6(=N_{\rm crit})$ case, orbital instabilities occur when $M_{\rm last}/M_{\rm init}\leq 0.92$ and $1.004\leq M_{\rm last}/M_{\rm init}$ except for $M_{\rm last}/M_{\rm init}=0.8$.
These indicate that the planets in resonant chains are less stable as their number increases even in $N\leq N_{\rm crit}$, and the planets cause orbital instability with a small mass change of 0.1\% -- 10\% when $N=N_{\rm crit}$.
The results also show that the amount of the mass gain to cause orbital instabilities is smaller than that of the mass loss.

In the $\mu$4$p$3 model, the transition from the stable resonant chain to the unstable one occurs at $N=7$. 
In one of the three simulation without any mass change, the planets with $N=7$ undergo orbital instability. 
We considered the planetary mass evolution for this initial condition.
We found that the planets with $N=7$ are stable only when they experience 0.1\% mass loss or 0.1\% mass gain. 
In simulations with more mass loss or mass gain, the systems cause orbital instabilities.

Knowing the orbital crossing time of planets in resonant chains that cause orbital instabilities would assist our judgment of whether planets in resonant chains are stable.
We compared the orbital crossing time in unstable cases with $t_{\rm cross,Z07}$ and $t_{\rm drag}$ (Equations (\ref{eq:t_Z}) and (\ref{eq:t_drag})) in the right panel of Figure \ref{Fig:num_Mml_t_cross_4_43}.
In the $\mu$4$p$3 model, $t_{\rm cross,Z07}$ is always shorter than $t_{\rm drag}$, and the disk gas depletion determines the orbital crossing time of planets that are not in resonant chains.
The orbital crossing time is almost equal to $t_{\rm drag}$ when $1.1<M_{\rm last}/M_{\rm init}$.
In contrast, the orbital crossing time is obviously longer than $t_{\rm drag}$ when $M_{\rm last}/M_{\rm init}<0.9$.
The orbital crossing time increases as $M_{\rm last}/M_{\rm init}$ decreases, while $t_{\rm drag}$ stays almost constant.
This feature is common regardless of the relationhip between $N$ and $N_{\rm crit}$.
The orbital crossing time in $N=6,$ 7, and 8 contains similar values to the $M_{\rm last}/M_{\rm init}$ range.
This reflects the longer orbital crossing time of smaller planets in the same resonances (Table \ref{table:models} and Appendix \ref{sect:reslt_pl_ap}).
While the relationship between $t_{\rm cross}$, $t_{\rm cross,Z07}$ and $t_{\rm drag}$ is different among our models (see Appendix \ref{sect:reslt_pl_ap}, Figures \ref{Fig:num_Mml_t_cross_5_65} -- \ref{Fig:num_Mml_t_cross_4_21}), the dependencies of $t_{\rm cross}$ on $M_{\rm last}/M_{\rm init}$ are similar.
When $0.9\leq M_{\rm last}/M_{\rm init}\leq1.1$, the orbital crossing times are sometimes significantly longer than $t_{\rm drag}$.
This mass evolution range is the transition from the stable resonant chain to the unstable one (see above $N=6$ cases).
The resonant effect partially works on the planets and their crossing time is longer than $t_{\rm drag}$.

In some models, planets are stable when $M_{\rm last}/M_{\rm init}= 0.5$ even in $N>N_{\rm crit}$ (e.g., $\mu5p5$ model in Appendix \ref{sect:mu5p5}).
This is because planets in resonant chains are more stable when their masses are small.
This means that there is a suitable range of planetary mass loss for planets to bring about orbital instability.
The unstable condition of the planets in resonant chains is $0.5 \lesssim M_{\rm last}/M_{\rm init}\lesssim 0.9$ when $N\simeq N_{\rm crit}$.
This corresponds to the situation that around 10\% -- 50\% masses of planets are composed of envelopes that are lost.
Such planetary mass loss occurs when planets are located at $\lesssim 0.1$~au \citep[e.g.,][]{Lopez+2012,Owen&Wu2013}.

\subsection{Dependence on the timescale of mass evolution}
\label{sect:t_ml}

\begin{figure}[hbt]
	\plotone{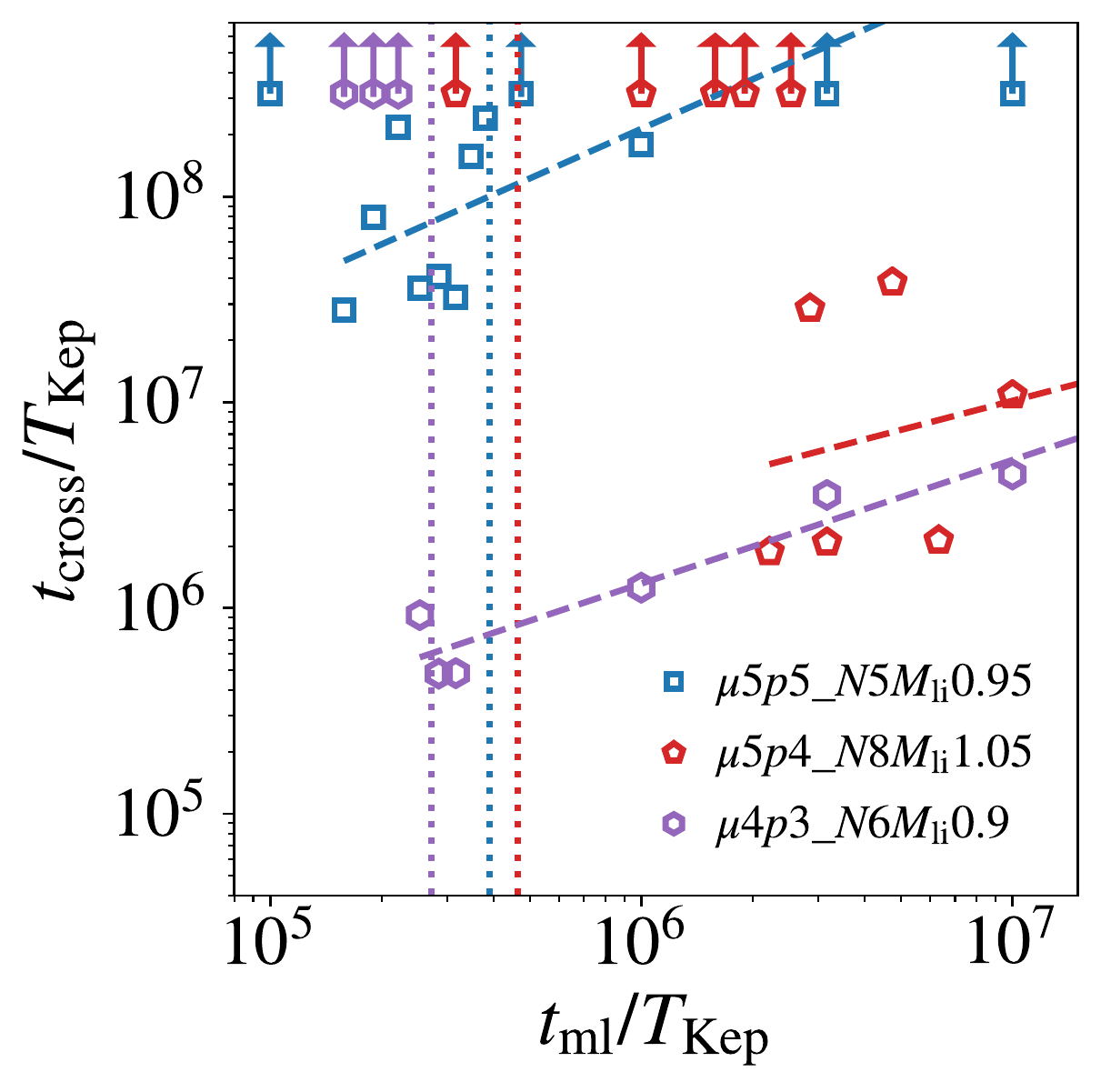}
	\caption{
		Dependence of the orbital crossing timescale on the timescale of the planet mass evolution ($t_{\rm ml}$).
		The vertical lines in the top-left section show $t_{\rm drag}$ of each setup.
		The dotted lines are the fitting lines ($\log{(t_{\rm cross}/T_{\rm Kep})}=C_{\rm ml1}\log{(t_{\rm ml}/T_{\rm Kep}) }+C_{\rm ml1}$) for each case: 
		$C_{\rm ml1}=0.80$ and $C_{\rm ml2}=3.5$ in the $\mu5p5$\_$N5M_{\rm li}0.95$ case;
		$C_{\rm ml1}=0.47$ and $C_{\rm ml2}=3.7$ in the $\mu5p4$\_$N8M_{\rm li}1.05$ case; 
		$C_{\rm ml1}=0.60$ and $C_{\rm ml2}=2.5$ in the $\mu4p3$\_$N6M_{\rm li}0.9 $ case.
	}
	\label{Fig:tml_tcross}
\end{figure}

While we set $t_{\rm ml}=10^4$~yr in our fiducial case, the timescale of the planetary mass evolution or stellar mass evolution would be longer than $10^4$~yr.
We examined the dependence of the orbital crossing time of planets in resonant chains on the timescale of the planetary mass evolution.
Figure \ref{Fig:tml_tcross} shows the crossing time as the function of the timescale of the planetary mass evolution.
For a clear presentation, we show orbital crossing times in three of five cases where we performed simulations.
The crossing time increases as the timescale of the planetary mass evolution increases.
In most cases, the crossing time is well expressed by the power-law function of $t_{\rm ml}$.
These dependencies are between 0.16 and 0.80, which are weaker than the linear relationship\footnote{
	The dependence of $t_{\rm cross}$ on $t_{\rm ml}$ in the other two cases are the followings:
	$C_{\rm ml1}=0.38$ and $C_{\rm ml2}=4.3$ in the $\mu5p5$\_$N6M_{\rm li}1.01$ case;
	$C_{\rm ml1}=0.16$ and $C_{\rm ml2}=5.6$ in the $\mu5p5$\_$N7M_{\rm li}0.95$ case.
}.
Even when planets in resonant chains experience longer timescale mass evolution, they cause orbital instabilities.

We found that the planets do not cause orbital instability when the mass evolution timescale is shorter than $\sim t_{\rm drag}$.
The planets in resonant chains are stable when $t_{\rm ml}/t_{\rm drag}$ is less than 0.41 -- 4.8
\footnote{
	In most cases, the boundary values of $t_{\rm ml}/t_{\rm drag}$ is less than 1. 
	The details are as follows:
	$t_{\rm ml}/t_{\rm drag}=0.41$ in the $\mu5p5$\_$N5M_{\rm li}0.95$ case;
	$t_{\rm ml}/t_{\rm drag}=0.58$ in the $\mu5p5$\_$N6M_{\rm li}1.01$ case;
	$t_{\rm ml}/t_{\rm drag}=0.41$ in the $\mu5p5$\_$N7M_{\rm li}0.95$ case;
	$t_{\rm ml}/t_{\rm drag}=4.8$ in the $\mu5p4$\_$N8M_{\rm li}1.05$ case; 
	$t_{\rm ml}/t_{\rm drag}=0.93$ in the $\mu4p3$\_$N6M_{\rm li}0.9 $ case.
}.
The longer mass evolution timescale is suitable to cause orbital instabilities of planets in resonant chains.
When $t_{\rm ml}<t_{\rm drag}$, planets are recaptured into the resonant chain due to gas drag.
To evaluate the gas drag effect, we performed simulations changing the onset time of planetary mass evolution in five cases.
We found that the onset time of planetary mass evolution does not affect the orbital crossing time even in the case that planetary mass evolution begins after $t_{\rm drag}$.

\subsection{Dependence on the fraction of planets with mass change}

\begin{figure}[htb]
	\plotone{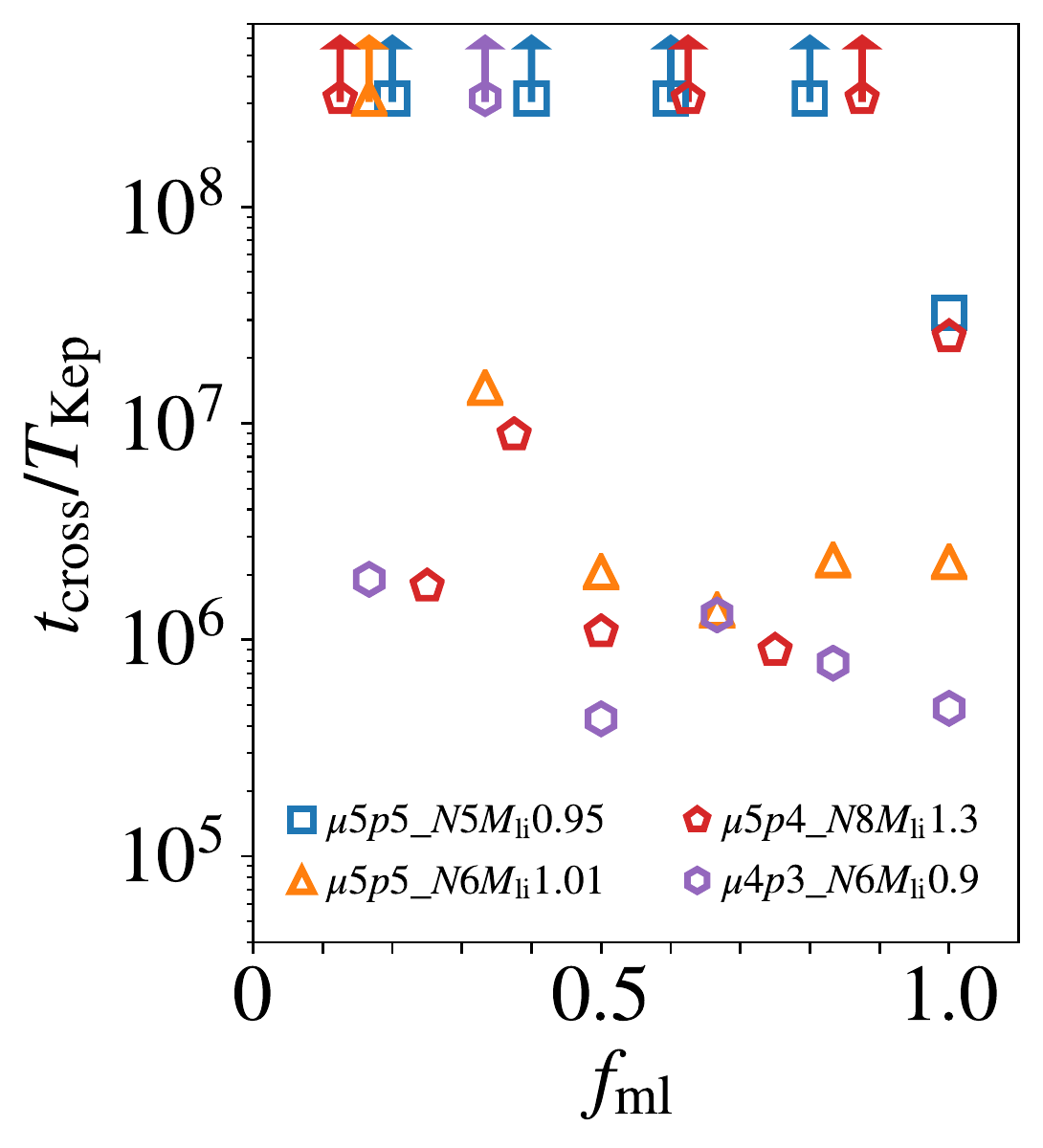}
	\caption{
		Dependence of the crossing timescale on the number fraction of planets that experience the mass evolution ($f_{\rm ml}$).
	}
	\label{Fig:Nin_tcross}
\end{figure}

The mass-loss rates of the planets in resonant chains are not uniform since the inner planets receive the stronger incident flux of the stellar radiation.
Some inner planets would lose their mass, while the other outer planets will not lose theirs.
We simulated this situation considering the number fraction of the planets that experience the mass evolution ($f_{\rm ml}$).

Figure \ref{Fig:Nin_tcross} shows the dependence of the crossing time on $f_{\rm ml}$ in four of five cases where we performed simulations.
Although the behavior of the crossing time on $f_{\rm ml}$ is not straightforward, we found several trends. 
We found that the stability of most systems does not depend on $f_{\rm ml}$ for $f_{\rm ml}>0.5$. 
In these systems, planets cause orbital instabilities when the inner half of them experience mass losses.
We also noticed that the resonant chain can be destabilized even when $f_{\rm ml} < 0.2$. 
This means that the resonant chain can be broken even when only the innermost planet undergoes the atmospheric loss. 
It is important to note, however, that most cases are stable when $f_{\rm ml} < 0.2$ in Figure \ref{Fig:Nin_tcross}.

\section{Results for the stellar mass evolution}\label{sect:result_st}

\subsection{Typical evolution}

\begin{figure}[ht]
	\plotone{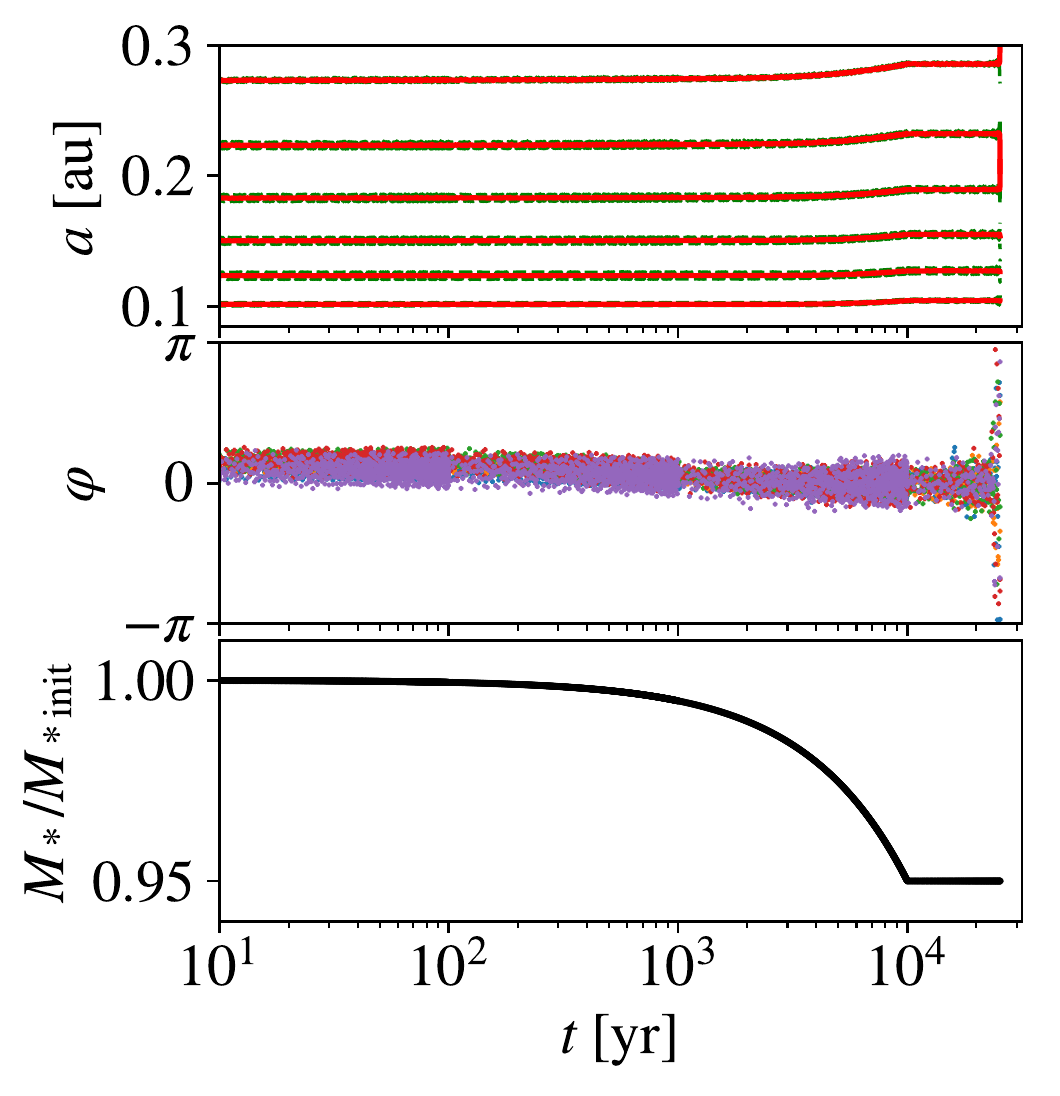}
	\caption{
		The same as Figure \ref{Fig:t_vs_aMphi_typical} but for the $\mu4p3$\_$N6M_{\rm *li}0.95$ case where the star loses its mass in the first $10^4$~yr.
		The orbital crossing between the fourth and fifth innermost planets occurs at $2.5\times 10^4$ yr.
		}
	\label{Fig:t_vs_aMphi_typical_Ms}
\end{figure}

We then investigated whether the stellar mass loss induces the orbital instability of planets in resonant chains. 
As stated in Section \ref{sect:intro}, stars can lose their masses by $\sim$0.1\% -- 1\% in the first 1~Gyr.
Figure \ref{Fig:t_vs_aMphi_typical_Ms} shows the time evolution of planets in the $\mu4p3$\_$N6M_{\rm *li}0.95$ case, where six $10^{-4}M_{\odot}$ mass planets are in 4:3 resonances around the central star whose final mass is $0.95M_{\odot}$.
The stellar mass decreases exponentially in the first $10^4$~yr, which causes the expansions of semimajor axes of planets \citep[][]{Minton&Malhotra2007}.
Although the evolutions of semimajor axes are suppressed in the first $\sim 10^3$~yr due to the remnant gas, the change in semimajor axes after $t\sim10^3$ years is evident.
We found that the libration widths of their resonant angles begin to increase at $2.3\times 10^4$~yr, and they begin circulations at $2.4\times 10^4$~yr.
As their eccentricities increase, the fourth and fifth innermost planets cause orbital crossing at $2.5\times 10^4$~yr.
Hence, we found that the stellar mass loss can also destabilize the resonant chain.
The orbital crossing time is longer than $t_{\rm cross,Z07}=95$~yr and $t_{\rm drag}=4.6\times 10^3$~yr.
The behavior of the orbital crossing time with the stellar mass loss is similar to that with the planetary mass loss.

The stellar mass evolution causes the expansion of semimajor axes. 
Both stellar mass and semimajor axes affect the Kepler times of planets.
In this section, orbital crossing times are not normalized by the Kepler time of the innermost planet.

\subsection{Dependence on the amplitude of the stellar mass evolution}

\begin{figure*}[hbt]
	\plottwo{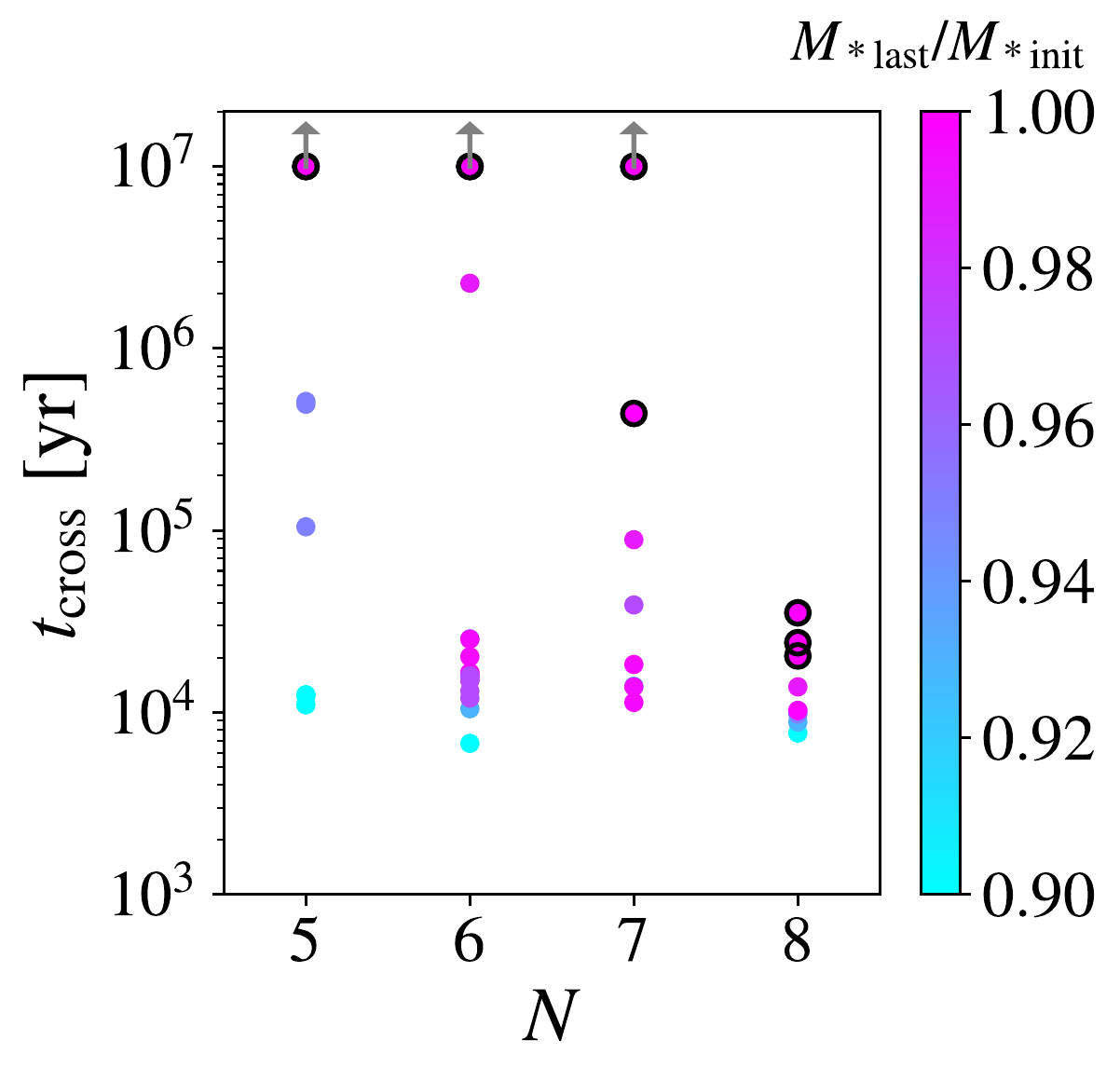}
			{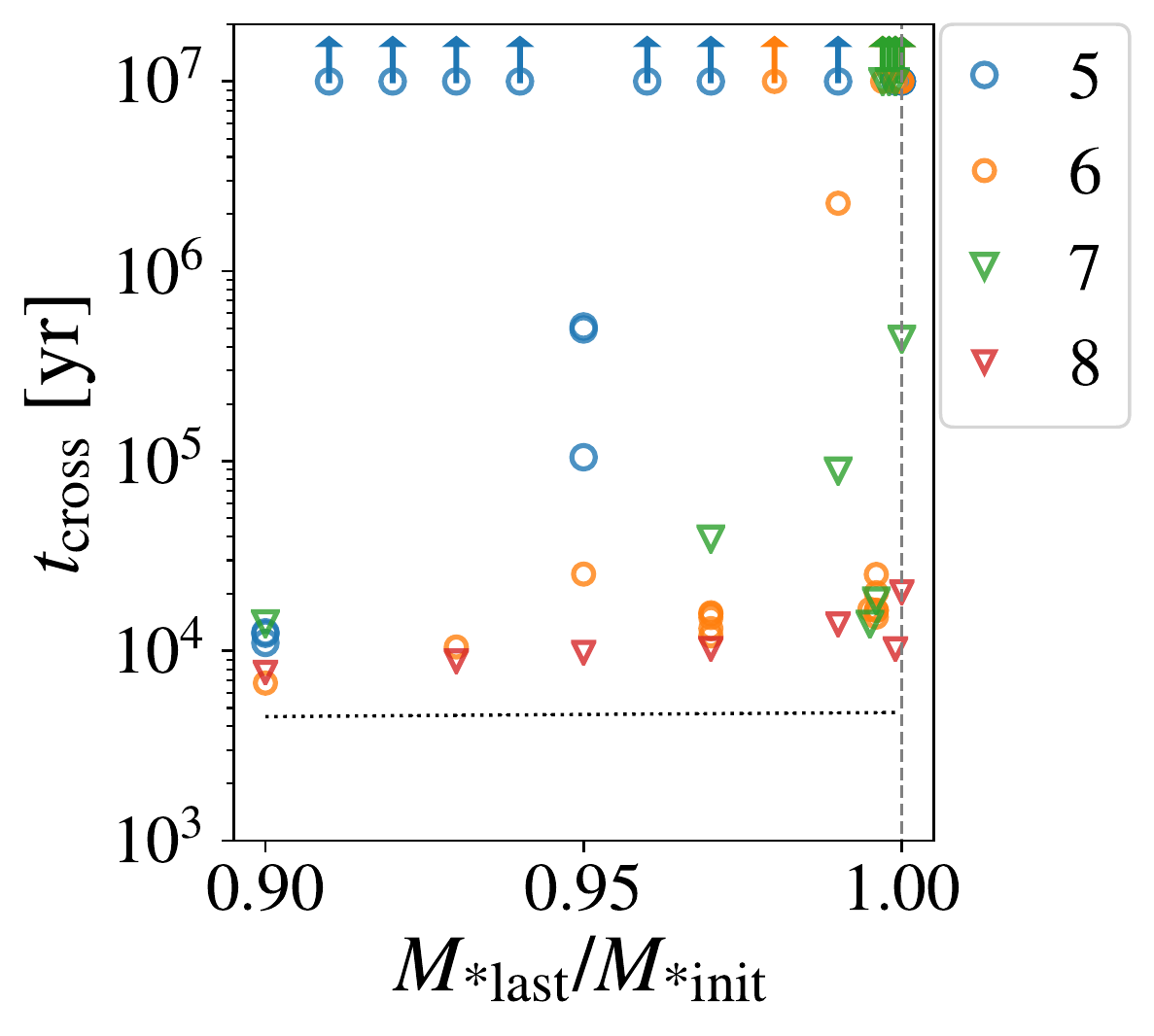}
	\caption{
		The same as Figure \ref{Fig:num_Mml_t_cross_4_43} but for the orbital crossing time as a function of $M_{\rm *last}/M_{\rm *init}$ in the $\mu$4$p$3 model.
		In the right panel, the vertical dashed line shows $M_{\rm *last}/M_{\rm *init}=1.0$, and the dotted line shows $t_{\rm drag}$.
		Note that $t_{\rm cross,Z07}$ does not appear in this panel since $t_{\rm cross,Z07}$ is shorter than $10^3$~yr.
	}
	\label{Fig:Mml_t_cross_4_43_Ms}
\end{figure*}

We then looked at the dependence of parameters on the orbit crossing time.
Figure \ref{Fig:Mml_t_cross_4_43_Ms} shows the orbital crossing time as the functions of $N$ (the left panel) and $M_{\rm *last}/M_{\rm *init}$ (the right panel) in the $\mu$4$p$3 model (see Appendix \ref{sect:reslt_st_ap} for models $\mu$5$p$5 and $\mu$5$p$4).
Similar to the planetary mass evolution, planets cause orbital instabilities when stars lose their masses.
Compared with the planet mass loss simulations (Figure \ref{Fig:num_Mml_t_cross_4_43}), the smaller amount of mass loss leads to  instability in the system. 
For the $N=5$ cases, we found that five planets are responsible for the orbital instability when $M_{\rm *last}/M_{\rm *init}=0.95$ and 0.90.
In the $N=6$ cases, orbital instabilities occur when $M_{\rm *last}/M_{\rm *init}\leq 0.996$ except for $M_{\rm *last}/M_{\rm *init}=0.98$.
Therefore, only 0.4\% of the mass change due to stellar mass loss can destabilize the resonant chain.
The stellar mass-loss rates that observations suggested \citep[][]{Wood+2002} are enough to bring about the orbital instabilities of planets, especially in $N=N_{\rm crit}$ cases.

\section{Conclusions}\label{sect:conclusion}

We investigated the orbital stability of planets in resonant chains including the mass evolution of planets or stars. 
We performed $N$-body simulations in six models for calculations in the planetary mass evolution and three models in the stellar mass evolution. 
Through these calculations, we obtained the orbital crossing times of planets in resonant chains. 
The features of these orbital crossing times are summarized as follows.
\begin{enumerate}
	\item 
	When the mass evolution of planets or stars is not considered, resonant chains are stable when $N\leq N_{\rm crit}$; however, they cause orbital instabilities when $N>N_{\rm crit}$.
	\item 
	When the planetary mass (either mass loss or mass gain) changes more than about 10\%, resonant systems with $N=N_{\rm crit}$ usually undergo orbit crossing and resonant chains can be broken. 
	Even when the mass change is small ($\sim$1\%), systems with closer resonances can undergo orbital instability.
	In other words, the critical number for orbital instability decreases by one or two.
	\item 
	When the amount of mass change is larger, resonant systems with $N<N_{\rm crit}$ can be destabilized. \item
	Systems in which all planets undergo mass evolution are more vulnerable to orbital instability than systems in which only a small fraction of planets exhibit mass change. 
	It is important to note, however, that, depending on resonant configurations, resonant chains can also be destabilized for systems in which only one or two planets lose their masses.
	\item 
	The stellar mass evolution can also induce resonant breaking. 
	The system can be destabilized even when the star loses only a minimal amount of mass ($<1$\%), which is plausible based on the stellar evolution.
\end{enumerate}
The results of this paper provide interesting insights. 
One reason for this is that, although the planetary mass loss can stabilize the system, resonant systems are destabilized due to the mass loss. 
In addition, other studies may draw on the results of this research.
For example, in \cite{Izidoro+2017}, the fraction of resonant systems that undergo orbital instabilities is inconsistent with the observed super-Earth systems: more fraction of resonant systems are formed than that of observed systems; the formed resonant systems tend to have more planets than the observed systems.
By incorporating mass evolutions into $N$-body simulations, more resonant systems would cause orbital instabilities and the observed systems would be reproduced more naturally. 

We can also discuss the origin of the observed systems in resonant chains.
Most of the observed planets in resonant chains are located at $\lesssim 0.1$~au \citep[][]{Mills+2016, MacDonald+2016,Jontof-Hutter+2016,Gillon+2017}.
These planets received strong stellar radiation, which causes planetary mass loss.
We suggest two scenarios why these planets stay resonant orbits. 
One is that the number of planets is less than $N_{\rm crit}-2$ of the resonant chains. 
In this case, planets do not cause orbital instabilities even if they lost $\gtrsim 10$~\% of their mass. 
The other one is that these planets did not experience mass loss since they did not have massive primordial atmospheres \citep[][]{Hori&Ogihara2020}.

\acknowledgments

We thank Doug Lin, Gabriele Pichierri, Yasunori Hori, Shinsuke Takasao, and Munehito Shoda for fruitful discussion.
We thank the anonymous referee for constructive comments that helped us to improve the manuscript.
This work was achieved using the grant of NAOJ Visiting Joint Research supported by the Research Coordination Committee, National Astronomical Observatory of Japan (NAOJ), National Institutes of Natural Sciences (NINS).
Numerical simulations were carried out on the PC cluster at the Center for Computational Astrophysics, National Astronomical Observatory of Japan and in Academia Sinica Institute for Astronomy and Astrophysics (ASIAA).

\appendix

\section{Stabilization due to the disk gas}\label{sect:stabilization}

In this section, we estimate the orbital crossing time of planets in a depleting gas disk.
These planets are stabilized by eccentricity damping emanating from the disk gas \citep[][]{Iwasaki+2001, Iwasaki+2002}.
As the gas depletes, the timescale of eccentricity damping ($t_e$) becomes longer.
When $t_e$ becomes longer than the orbital crossing time of planets, their eccentricities are no longer damped, and the orbital crossing time becomes equal to those in the gas-free condition.

First, we estimated the orbital crossing timescale of the planets that were not present in resonant orbits, which is a well-studied area of research \citep[e.g.,][]{Chambers+1996, Yoshinaga+1999, Zhou+2007, Smith&Lissauer2009, Pu&Wu2015, Rice+2018}.
We used the empirical fitting formula in \cite{Zhou+2007} to estimate the orbital crossing timescale, which includes the dependence of the planet--star mass ratio.
In the empirical equation, the orbital crossing timescale ($t_{\rm cross,Z07}$) is described as
\begin{eqnarray}
	\log{\left( \frac{t_{\rm cross,Z07}}{ T_{\rm Kep} } \right)} 
	&=& A + B\log{\left( \frac{ \Delta a/r_{\rm H,init} }{ 2.3} \right)} 
	- \frac{B}{3} 
	\log{\left( \frac{\mu}{\mu_{\rm init}} \right) }
	,\nonumber\\
	A &=& -2 + {\tilde e} -0.27 \log{\mu}, \nonumber \\
	B &=& (18.7 +1.1\log{\mu} ) - (16.8 + 1.2\log{\mu}) {\tilde e},
	\nonumber \\
	{\tilde e} &=& \frac{e_0/h}{ 0.5\Delta a/r_{\rm H} },
	\label{eq:t_Z}
\end{eqnarray}
where $e_0$ is the initial eccentricities of the planets, $h$ is the reduced Hill radius of the planets ($h=r_{\rm H}/a$).
The mass evolutions of planets and stars affect $t_{\rm cross,Z07}$ via $\log{\mu}$.
The orbital crossing timescale becomes shorter as $\mu$ increases.

Now, we derive the stabilization timescale of the depleting disk gas.
Considering $t_e$ is equal to $t_{\rm cross,Z07}$, the timescale of the stabilization due to $e$-damping ($t_{\rm drag}$) is estimated as 
\begin{eqnarray}
	t_{\rm drag} = 2.3 t_{\rm dep} \log{ \left( \frac{ t_{\rm cross,Z07} }{ t_{e}(f_{\rm g}=1) } \right) },
	\label{eq:t_drag}
\end{eqnarray}
which is $\sim 10 t_{\rm dep}$
\footnote{
	In Equation (\ref{eq:t_drag}), we transformed the base of the logarithm from $e$ to 10, which makes the factor of 2.3.
}.
After $t_{\rm drag}$ has passed, eccentricity damping is no longer effective, and the planets not present in resonances would become unstable in $t_{\rm cross,Z07}$.
The maximum value between $t_{\rm cross,Z07}$ and $t_{\rm drag}$ approximately gives the orbital crossing timescale of non-resonant planets.
In the estimation of $t_{\rm drag}$, we substitute $M=M_{\rm last}$ and $M_* = M_{\rm *last}$.

\section{Results of each planetary mass evolution model}\label{sect:reslt_pl_ap}

We performed simulations in models $\mu$5$p$5, $\mu$5$p$4, $\mu$5$p$3, $\mu$4$p$2, and $\mu$4$p$1, including the planetary mass evolution.
In this section, we show the results of these models, which exhibit a similar tendency to the results in the $\mu$4$p$3 model.
We found that more mass gain is needed to bring about orbital instabilities when the orbital separations are larger or the number of planets is smaller.
Similarly, in mass loss cases, the mass change range where planets cause orbital instabilities is smaller as the orbital separations are larger or the number of planets is smaller.

\subsection{$\mu$5$p$5 model}\label{sect:mu5p5}

\begin{figure*}[ht]
	\plottwo{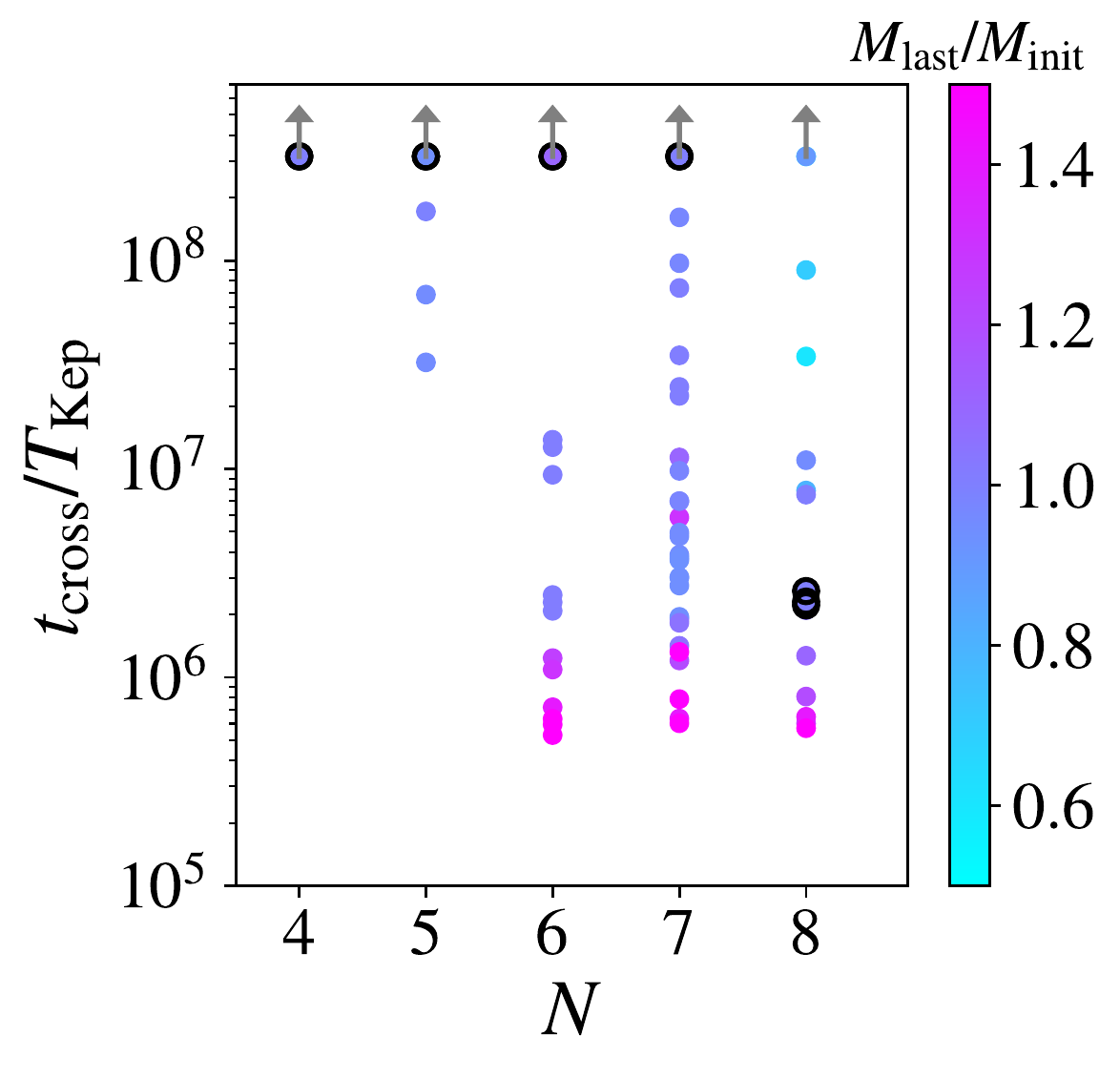}
			{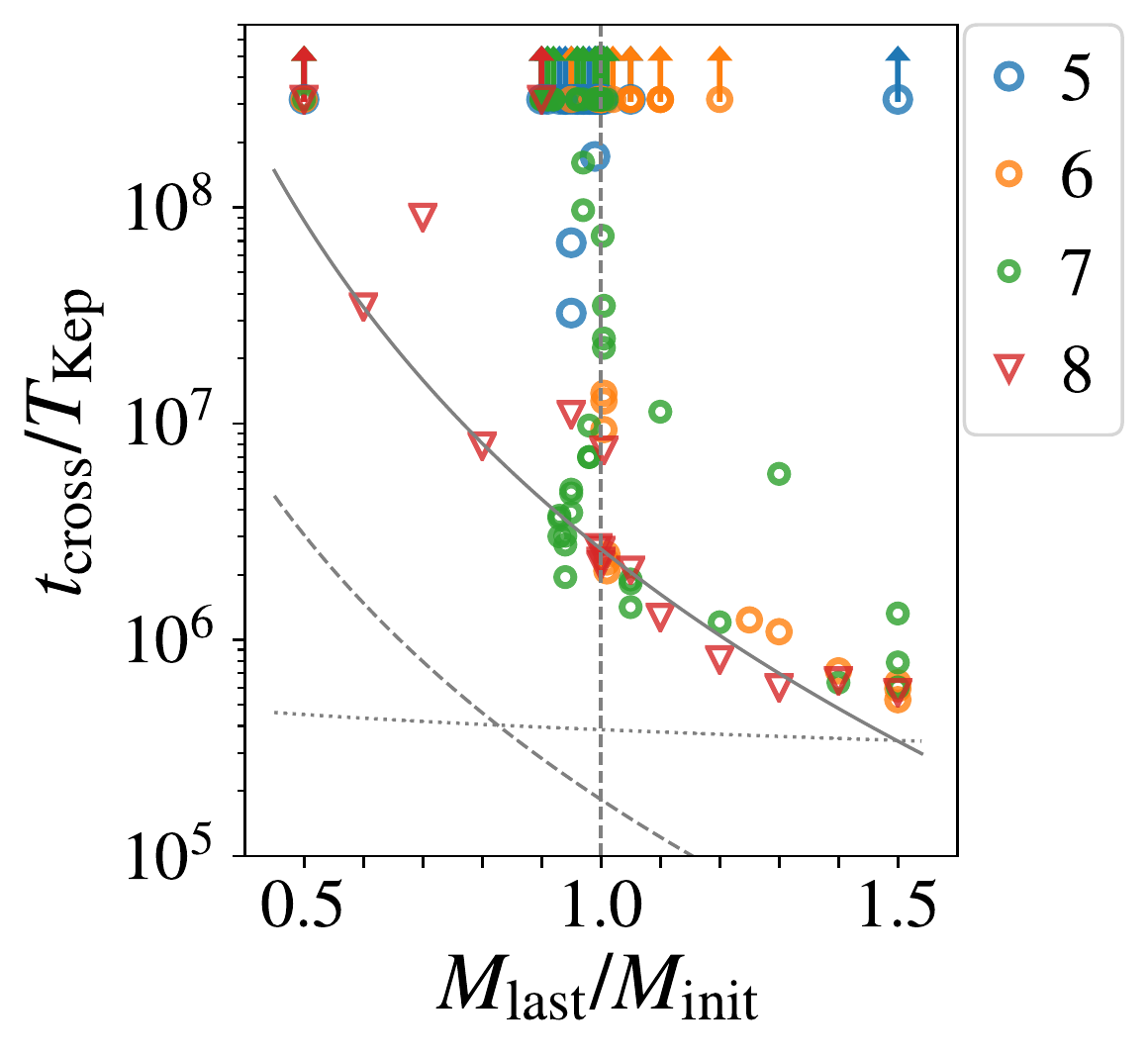}
	\caption{
		The same as Figure \ref{Fig:num_Mml_t_cross_4_43} but for the orbital crossing time of planets that initially have $10^{-5}M_*$ in 6:5 resonances (Model $\mu$5$p$5).
		The solid fitting line for local short crossing times is
		$\log{(t_{\rm cross}/T_{\rm Kep})} = -5.0\log{(M_{\rm last}/M_{\rm init})} + 6.4$.
	}
	\label{Fig:num_Mml_t_cross_5_65}
\end{figure*}

The orbital crossing time in the $\mu$5$p$5 model is shown as functions of $N$ and $M_{\rm last}/M_{\rm init}$ in Figure \ref{Fig:num_Mml_t_cross_5_65}.
In this case, the critical number is $N_{\rm crit}=7$.
In $N=7$ cases, planets are stable in $0.99 \leq M_{\rm last}/M_{\rm init}\leq 1.002$. 
The transition from the stable to the unstable resonances is between $M_{\rm last}/M_{\rm init}=1.003$ and 1.01 in mass gain cases.
Planets cause orbital instabilities when $1.05 \leq M_{\rm last}/M_{\rm init}$.
In the mass loss cases, while planets cause orbital instabilities in $0.93\leq M_{\rm last}/M_{\rm init}\leq 0.98$ except for $M_{\rm last}/M_{\rm init}=0.96$, they are stable in $0.92\leq M_{\rm last}/M_{\rm init}$.
This suggests that there are three regimes for orbital crossing times: planets do not cause orbital instabilities in slight mass loss cases; planets cause orbital instabilities in moderate mass loss cases; planets are stable in resonant chains in large mass loss cases. 
The orbital crossing time of planets in large separation resonant chains is longer \citep[][]{Matsumoto+2012}; that is, the orbital crossing time is longer as planets are smaller in the same resonant chains (Table \ref{table:models} and Section \ref{sect:mu5p3}).
Moreover, planets are stable in $M_{\rm last}/M_{\rm init}=0.5$ simulations even in $N=8$ cases.

\subsection{$\mu$5$p$4 model}

\begin{figure*}[ht]
	\plottwo{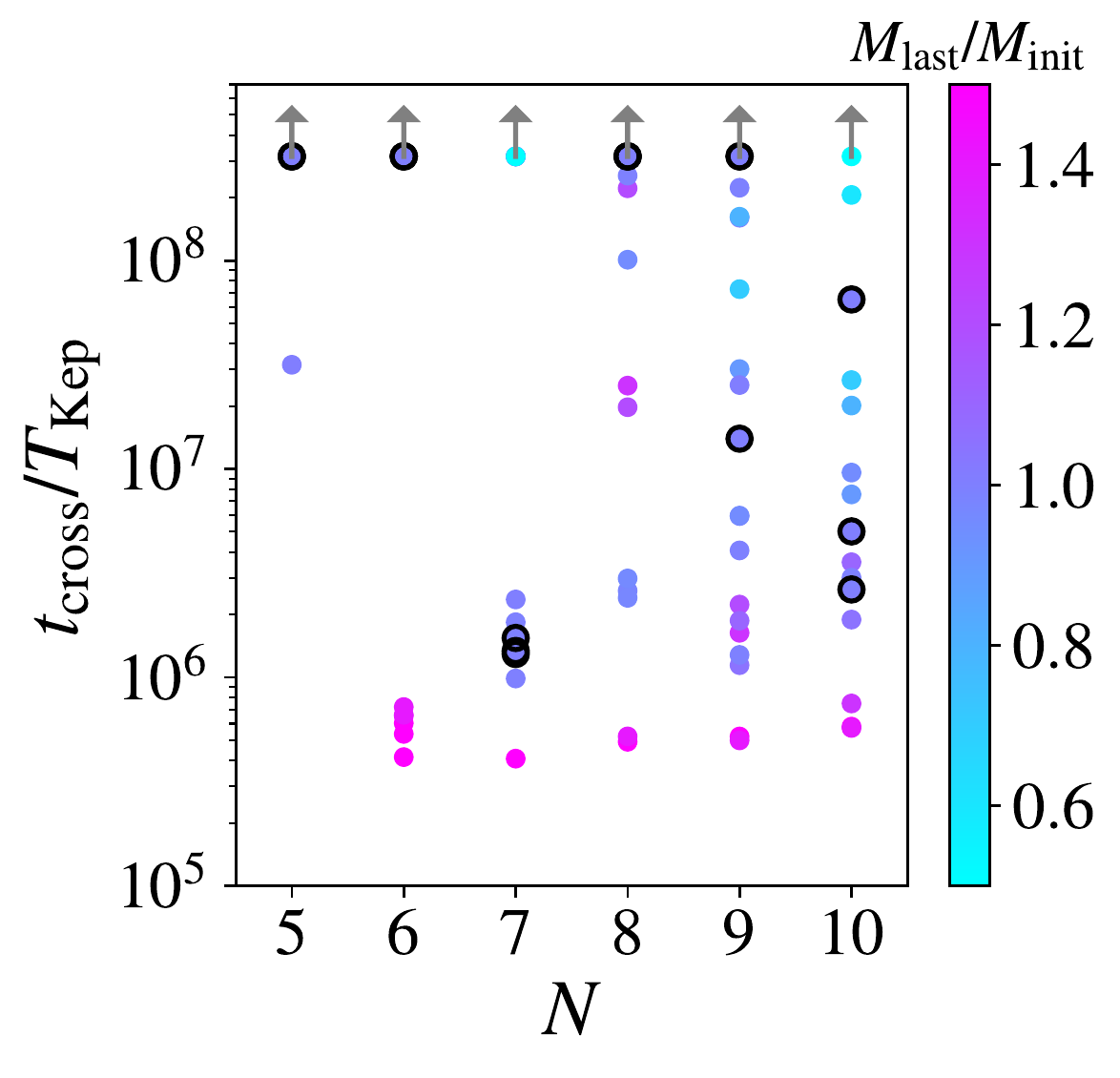}
			{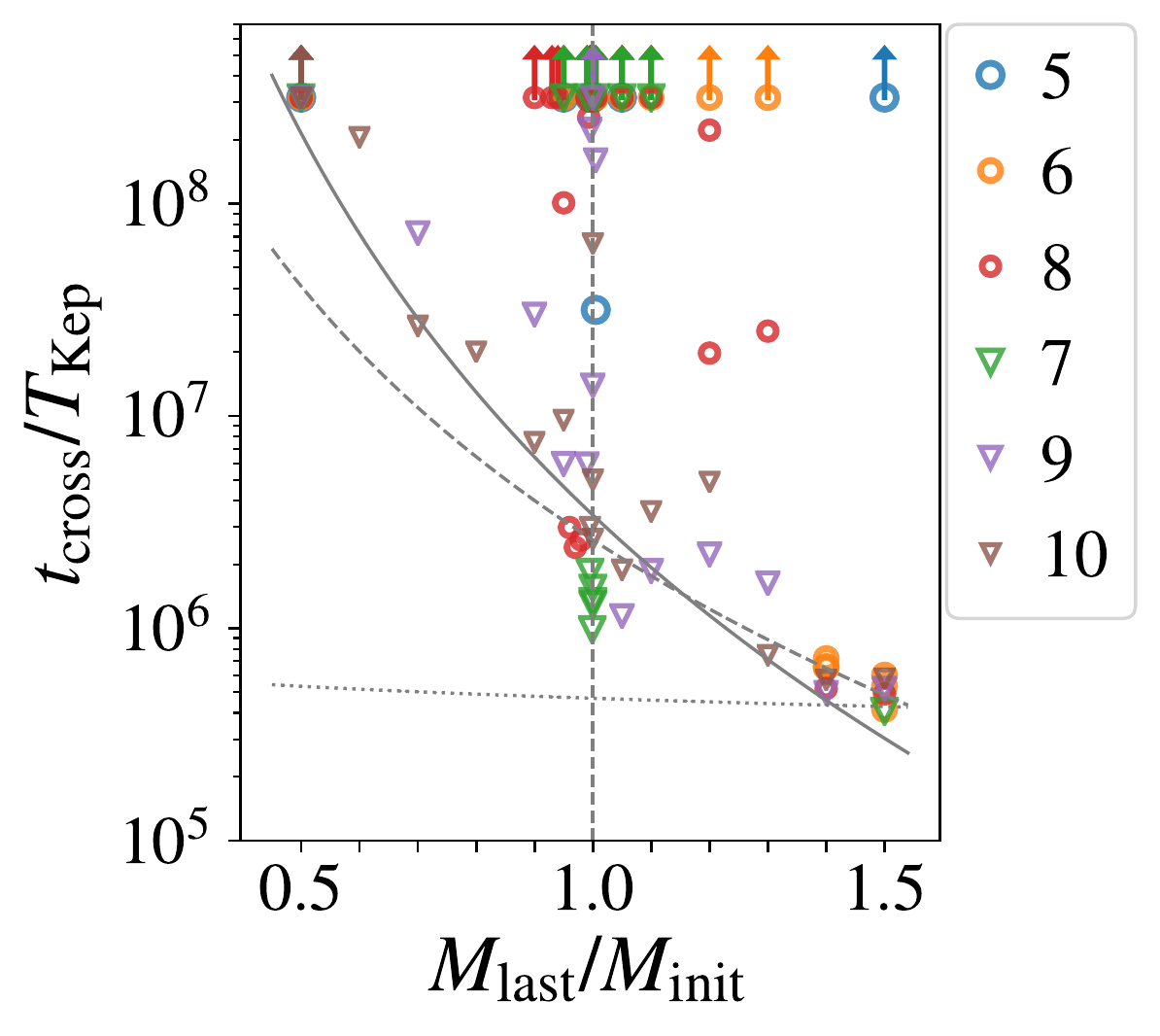}
	\caption{
		The same as Figure \ref{Fig:num_Mml_t_cross_4_43} but for the orbital crossing time of planets that initially have $10^{-5}M_*$ in 5:4 resonanes (Model $\mu$5$p$4).
		The solid fitting line for local short crossing times is
		$\log{(t_{\rm cross}/T_{\rm Kep})} = -6.0\log{(M_{\rm last}/M_{\rm init})} + 6.5$.
	}
	\label{Fig:num_Mml_t_cross_5_54}
\end{figure*}

The orbital crossing time in the $\mu$5$p$4 model is in Figure \ref{Fig:num_Mml_t_cross_5_54}.
In this case, the critical number is $N_{\rm crit}=6$.
The transition of the orbital stability of planets without mass loss is from $N=7$ to 9.
In $N=8$ cases and some $N=9$ simulations, planets are stable over $10^{8.5} T_{\rm Kep}$. 
In transition, slight mass loss leads planets toward stable orbits.
In $N=7$ cases, planets are stable in $ M_{\rm last}/M_{\rm init}\leq0.99$ and $1.005\leq M_{\rm last}/M_{\rm init}\leq1.1$.

The mass changes to bring about orbital instabilities are as follows.
In $N=6$ cases, planets cause orbital instabilities in $M_{\rm last}/M_{\rm init}\geq1.4$.
In $N=8$ cases, orbital instabilities occur in $M_{\rm last}/M_{\rm init}=0.993$, $0.95\leq M_{\rm last}/M_{\rm init}\leq 0.98$, and $M_{\rm last}/M_{\rm init}\leq 1.2$.

The orbital crossing time of planets in unstable resonant chains steeply increases than the line of $t_{\rm cross,Z07}$.
When $M_{\rm last}/M_{\rm init}=0.5$, the orbital crossing time is longer than $10^{8.5}T_{\rm Kep}$, even in $N=10$.

\subsection{$\mu$5$p$3 model}\label{sect:mu5p3}

\begin{figure*}[ht]
	\plottwo{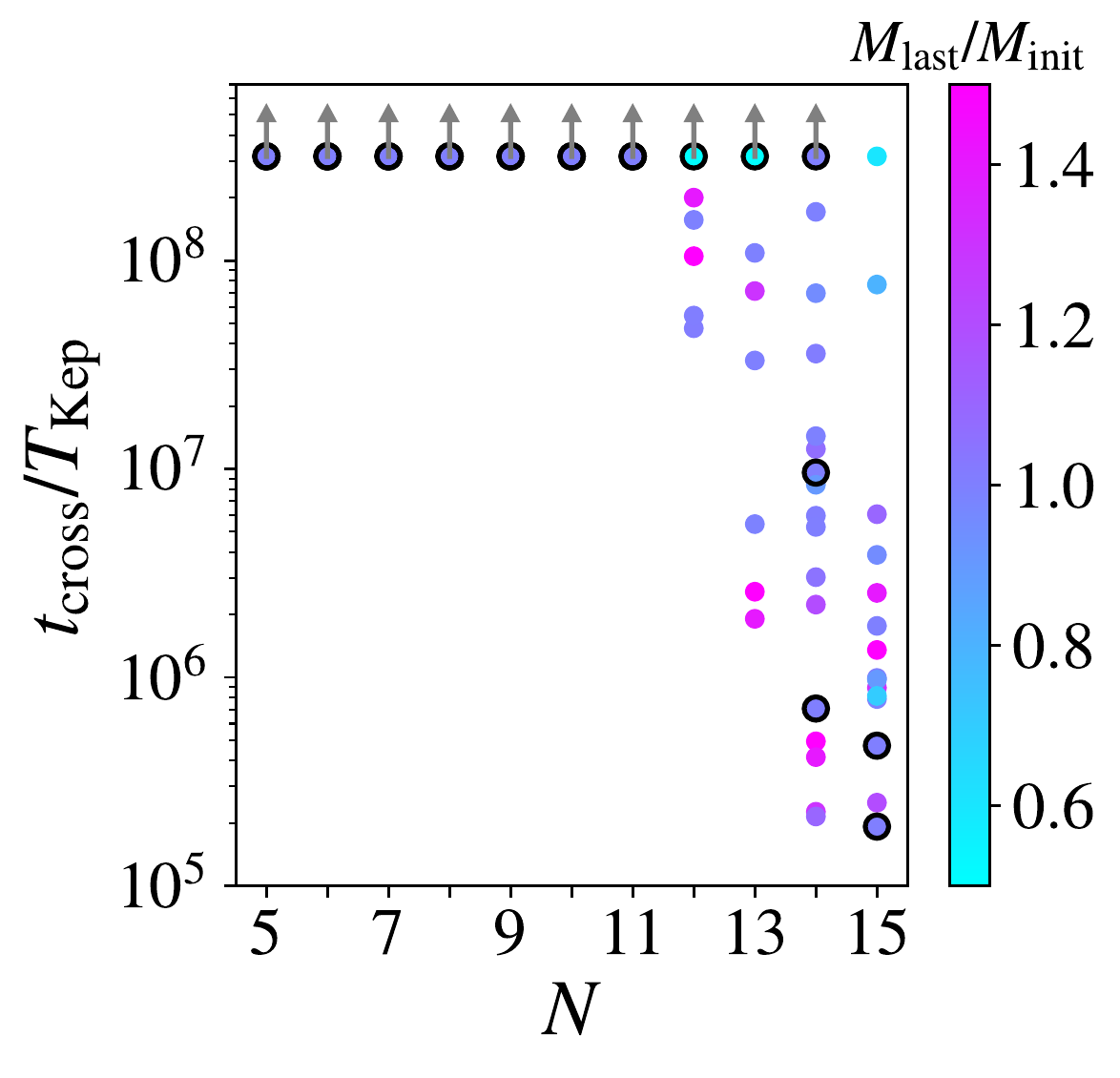}
			{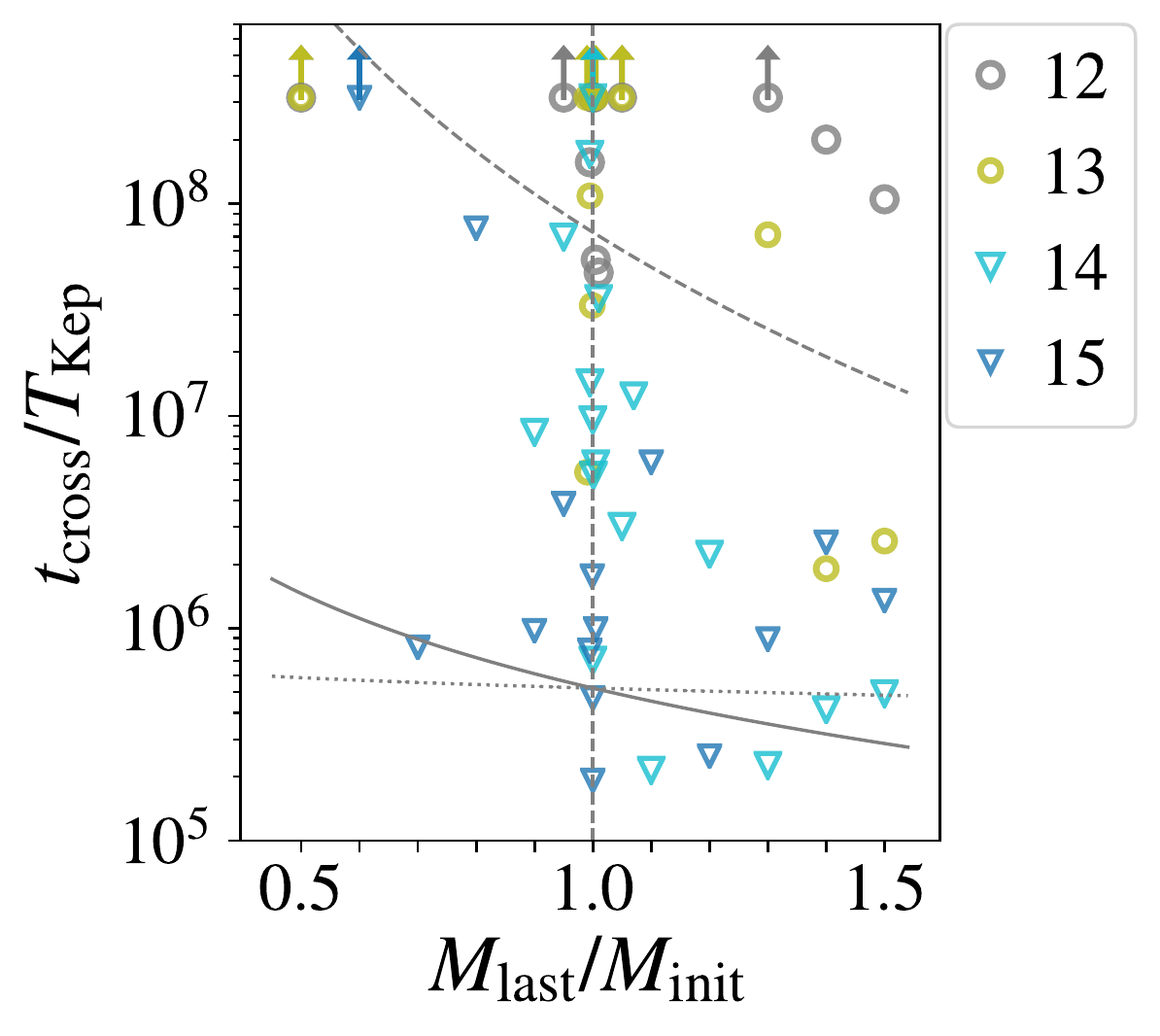}
	\caption{
		The same as Figure \ref{Fig:num_Mml_t_cross_4_43} but for the orbital crossing time of planets that initially have $10^{-5}M_*$ in 4:3 resonanes (Model $\mu$5$p$3).
		The solid fitting line for local short crossing times is
		$\log{(t_{\rm cross}/T_{\rm Kep})} = -1.5\log{(M_{\rm last}/M_{\rm init})} + 5.7$.
	}
	\label{Fig:num_Mml_t_cross_5_43}
\end{figure*}

The orbital crossing time in the $\mu$5$p$3 model is in Figure \ref{Fig:num_Mml_t_cross_5_43}.
The critical number is $N_{\rm crit}=13$, which is larger than the critical number in the $\mu$4$p$3 model ($N_{\rm crit}=6$).
Planets are more stable when their masses are small in the same resonances.
Dependencies of $N_{\rm crit}$ are understood by the orbital crossing time of planets not present in resonant orbits \citep[][]{Matsumoto+2012}.
The orbital crossing timescale is longer when planetary mass decreases because their mutual perturbations are weaker (Section \ref{sect:stabilization}).
The same applies to the orbital stability of planets in resonant chains.
In this model, planets do not cause orbital instbaility in $0.5\leq M_{\rm last}/M_{\rm init}\leq 1.5$ when $N\leq 11$.
 
The condition in which the planets cause orbital instabilities is as follows.
In $N=12$ cases, planets cause orbital instabilities in $M_{\rm last}/M_{\rm init}=0.995$, $1.005\leq M_{\rm last}/M_{\rm init}\leq1.01$, and $1.4\leq M_{\rm last}/M_{\rm init}$.
In $N=13$ cases, planets cause orbital instabilities in $0.991\leq M_{\rm last}/M_{\rm init}\leq0.999$, and $1.3\leq M_{\rm last}/M_{\rm init}$.
The orbital crossing time in the mass loss simulations tends to be longer than that in the mass gain simulations.
In large $\Delta a/r_{\rm H,init}$ models, planets tend to stay in resonant chains.
Planets are stable even in $N=15$ when $M_{\rm last}/M_{\rm init}\leq 0.6$.
It is important to note that the estimated orbital crossing time of planets not present in resonant chains ($t_{\rm cross,Z07}$) is longer than $10^{8.5}T_{\rm Kep}$ when $M_{\rm last}/M_{\rm init}\leq 0.6$.

\subsection{$\mu$4$p$2 model}

\begin{figure*}[ht]
	\plottwo{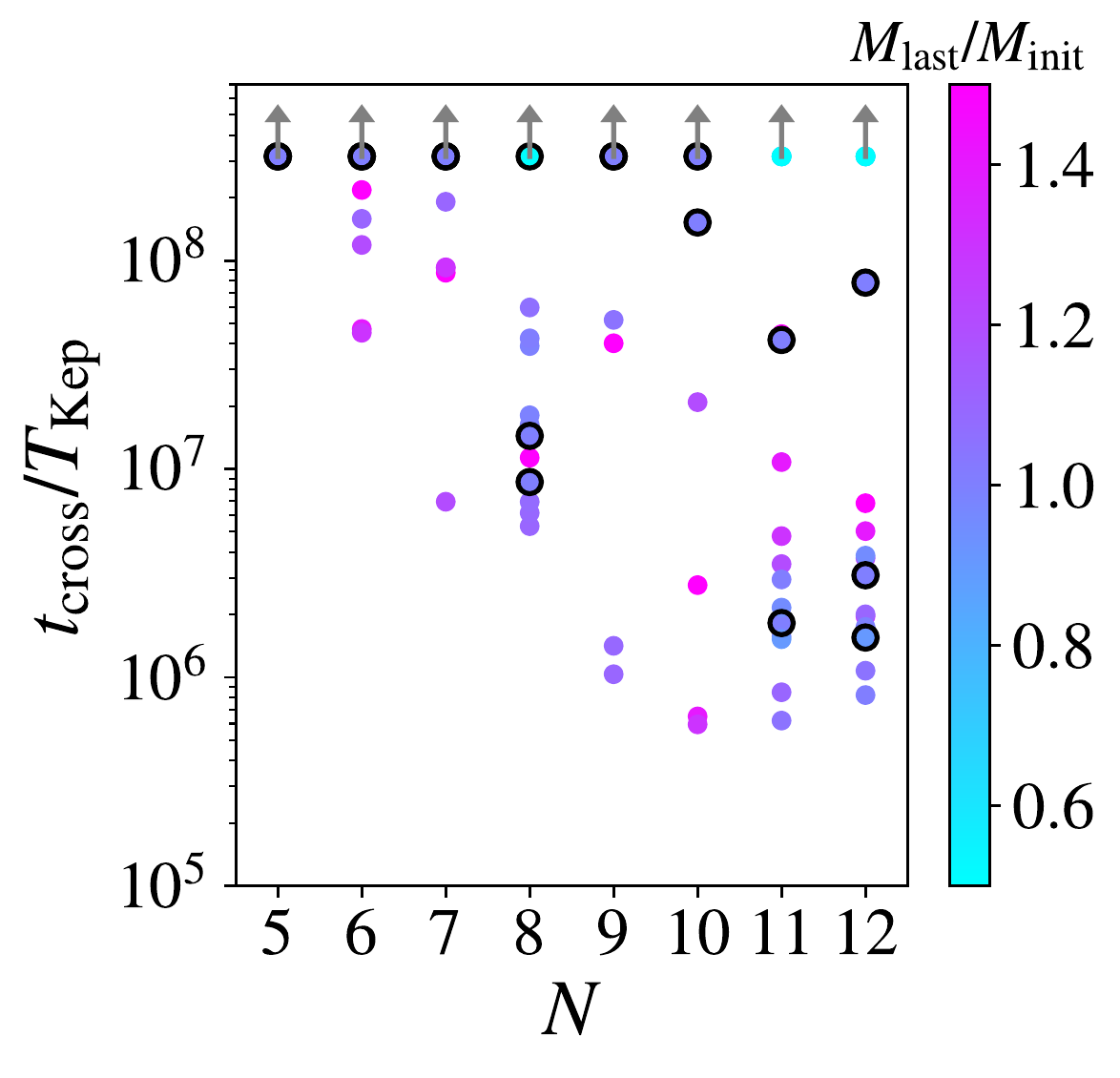}
			{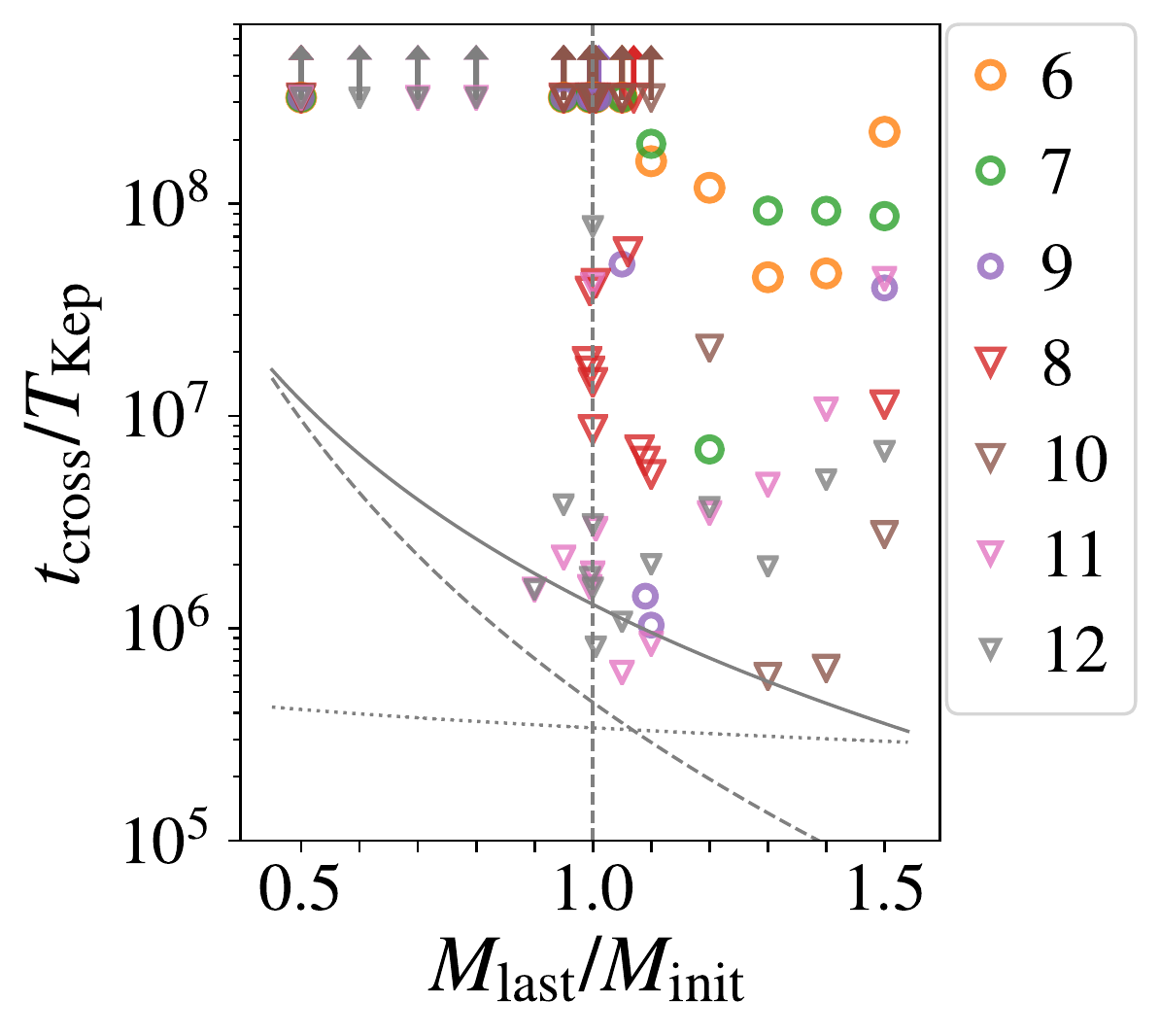}
	\caption{
		The same as Figure \ref{Fig:num_Mml_t_cross_4_43} but for the orbital crossing time of planets in 3:2 resonanes (Model $\mu$4$p$2).
		The solid fitting line for local short crossing times is
		$\log{(t_{\rm cross}/T_{\rm Kep})} = -3.2\log{(M_{\rm last}/M_{\rm init})} + 6.1$.
	}
	\label{Fig:num_Mml_t_cross_4_32}
\end{figure*}

The orbital crossing time in the $\mu$4$p$2 model is in Figure \ref{Fig:num_Mml_t_cross_4_32}.
The critical number is $N_{\rm crit}=7$.
The transition of the orbital stability of planets without mass loss is from $N=8$ to 10.
The orbital crossing time in this model shows a similar tendency to that in the $\mu$5$p$3 model.
The planets do not tend to cause orbital instabilities in the mass loss simulations but tend to bring about orbital instabilities in the mass gain simulations.
The planets cause orbital instabilities in the mass gain simulations when $1.1\leq M_{\rm last}/M_{\rm init}$ and $N=6$ while $N\leq N_{\rm crit}$ planets do not cause orbital instabilities in the mass loss simulations.
When $M_{\rm last}/M_{\rm init}\leq0.8$, planets are stable, even in $N=12$.

\subsection{$\mu$4$p$1 model}

\begin{figure*}[ht]
	\plottwo{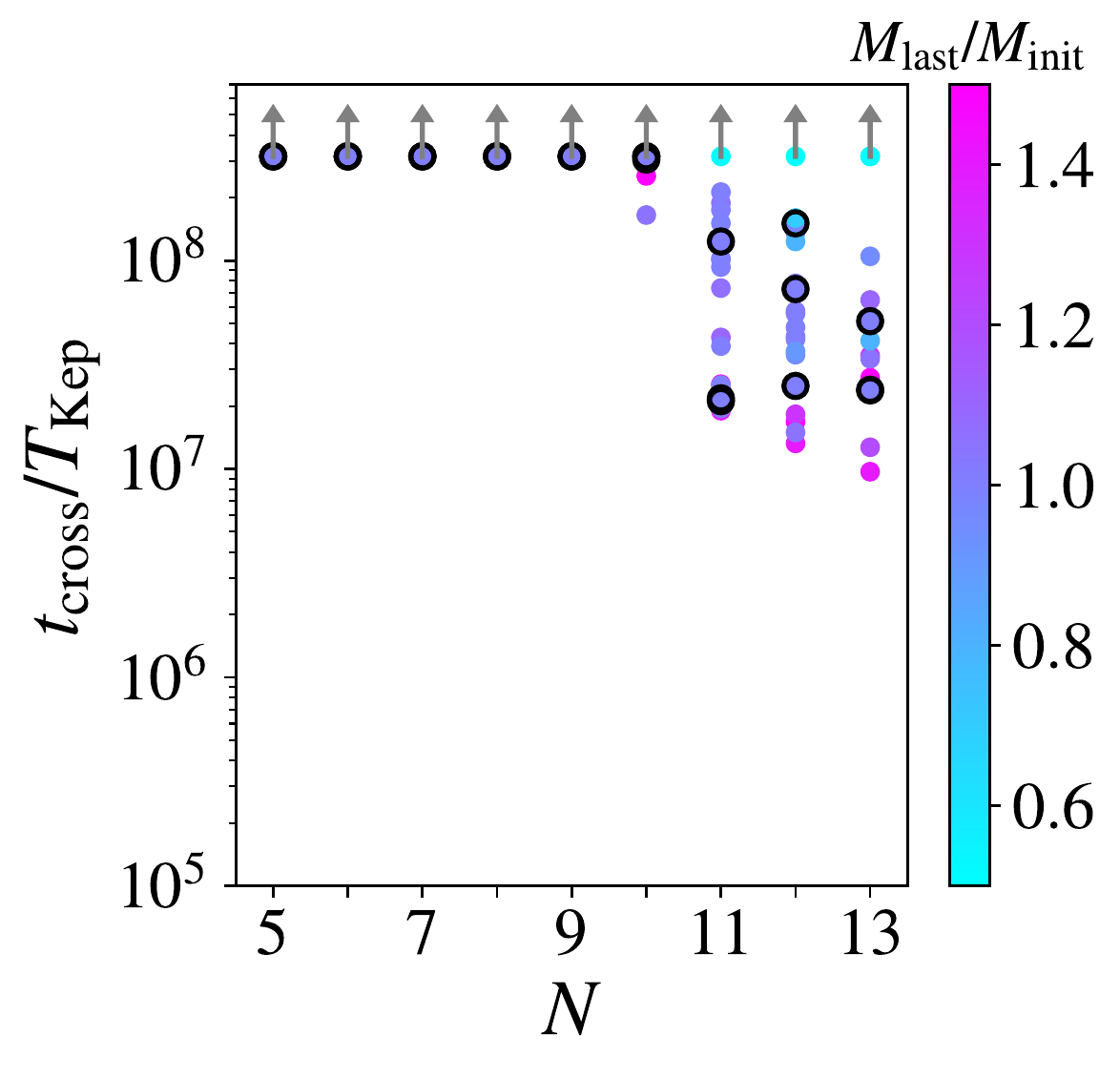}
			{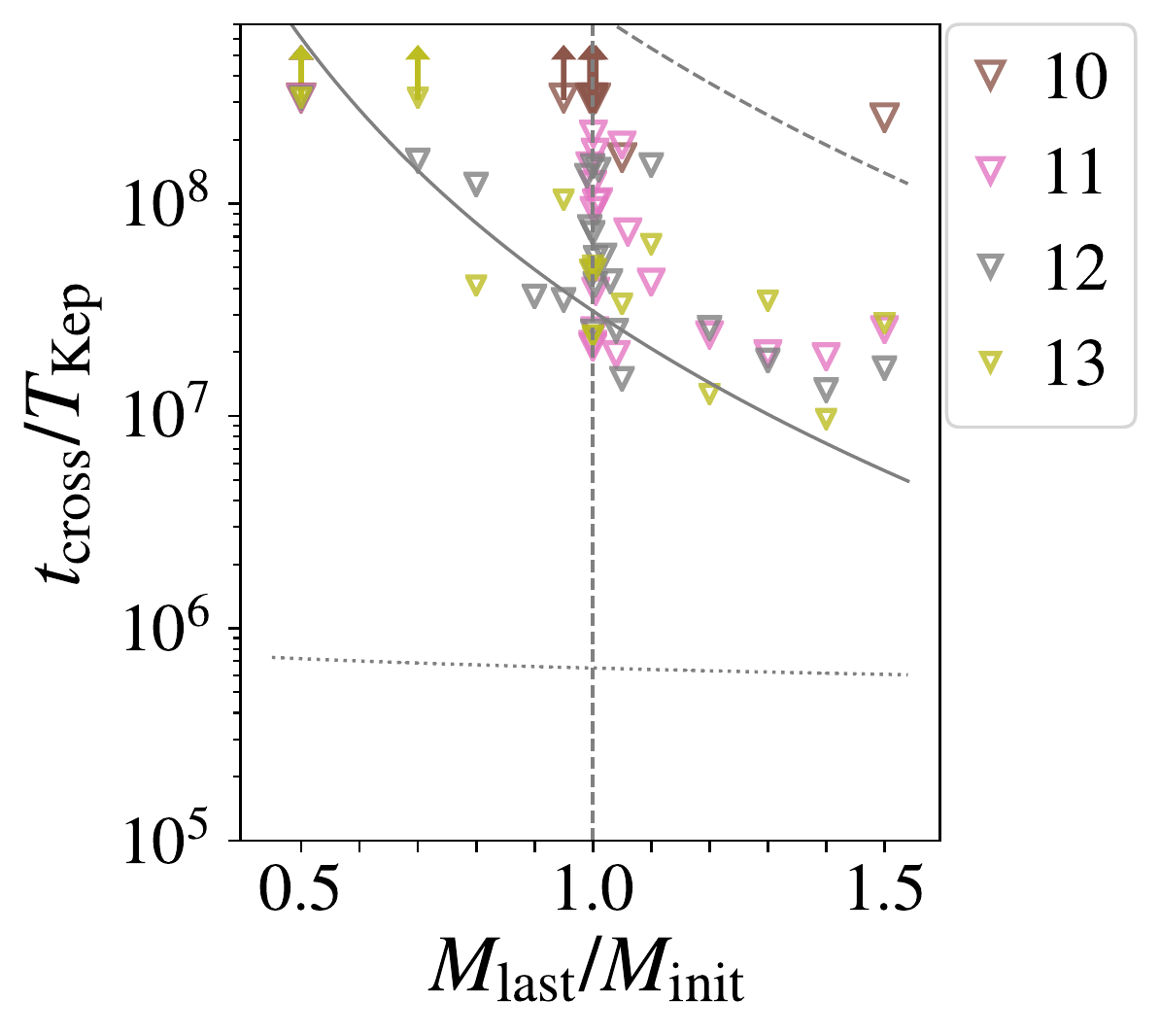}
	\caption{
		The same as Figure \ref{Fig:num_Mml_t_cross_4_43} but for the orbital crossing time of planets in 2:1 resonanes (Model $\mu$4$p$1).
		The solid fitting line for local short crossing times is
		$\log{(t_{\rm cross}/T_{\rm Kep})} = -4.3\log{(M_{\rm last}/M_{\rm init})} + 7.5$.
	}
	\label{Fig:num_Mml_t_cross_4_21}
\end{figure*}

The orbital crossing time in the $\mu$4$p$1 model is in Figure \ref{Fig:num_Mml_t_cross_4_21}.
The critical number is $N_{\rm crit}=9$.
The transition of the orbital stability of planets without mass loss is $N=10$.
In this model, planets do not cause orbital instabilities within $10^{8.5}T_{\rm Kep}$ in $N\leq9$, even if we consider the mass evolution in $0.5\leq M_{\rm last}/M_{\rm init}\leq 1.5$.
It is worth noting that $t_{\rm cross,Z07}$ is longer than $10^{8.5}T_{\rm Kep}$ in $M_{\rm last}/M_{\rm init}<1.25$.

\section{Stellar mass evolution results of each model}\label{sect:reslt_st_ap}

\begin{figure*}[ht]
	\plottwo{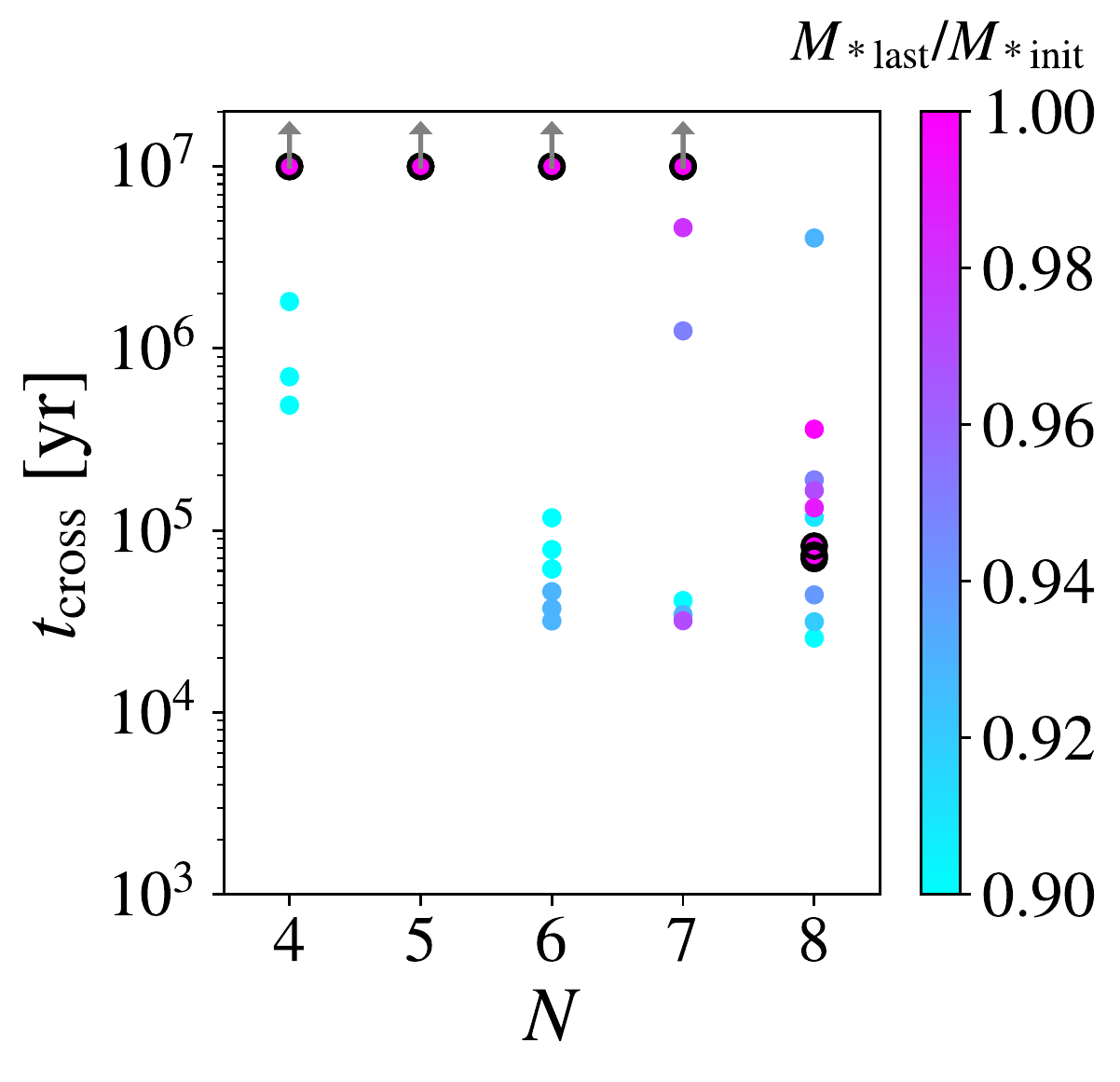}
			{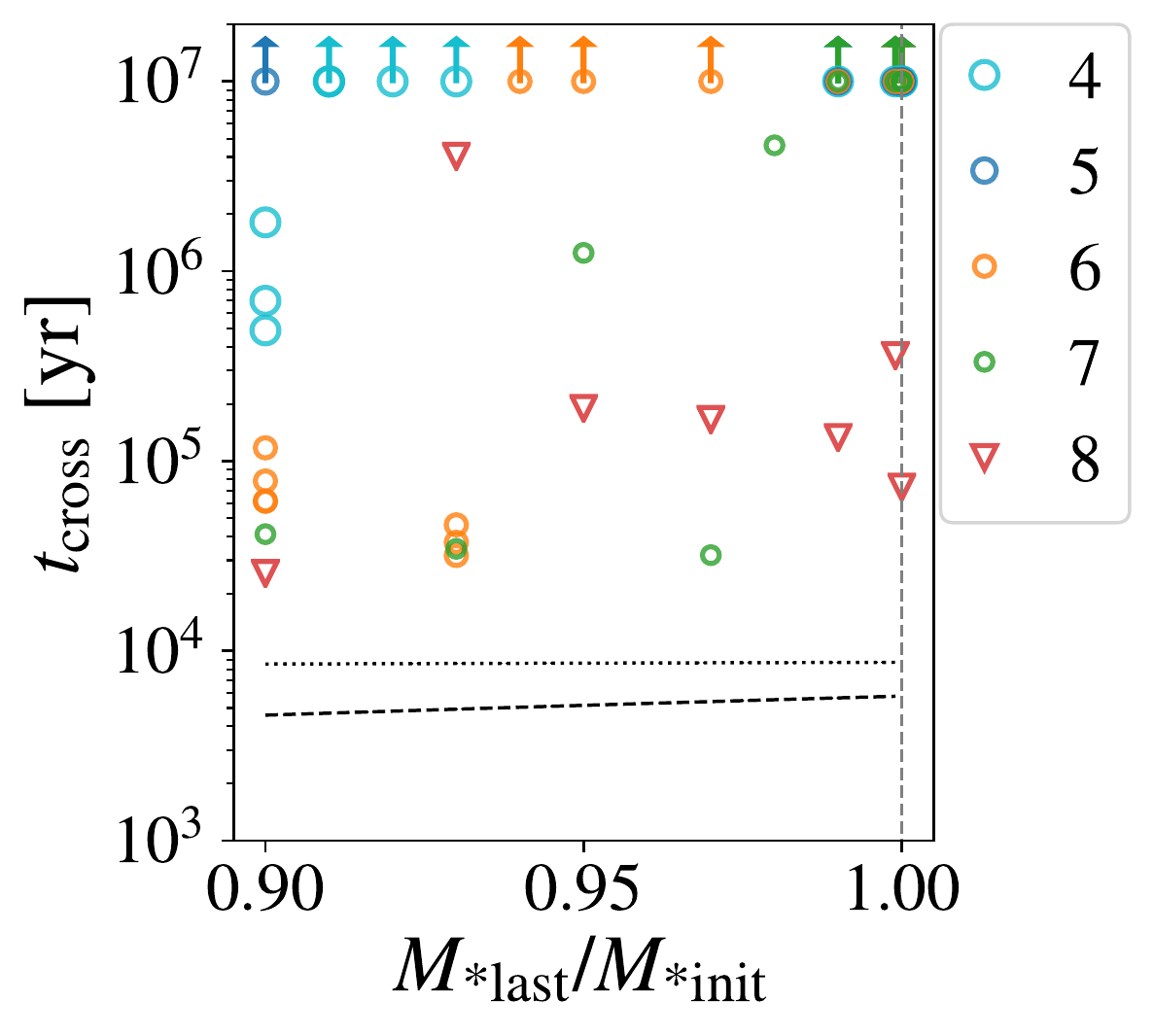}
	\caption{
		The same as Figure \ref{Fig:Mml_t_cross_4_43_Ms} but for the $\mu$5$p$5 model.
		}
	\label{Fig:Mml_t_cross_5_65_Ms}
\end{figure*}

\begin{figure*}[ht]
	\plottwo{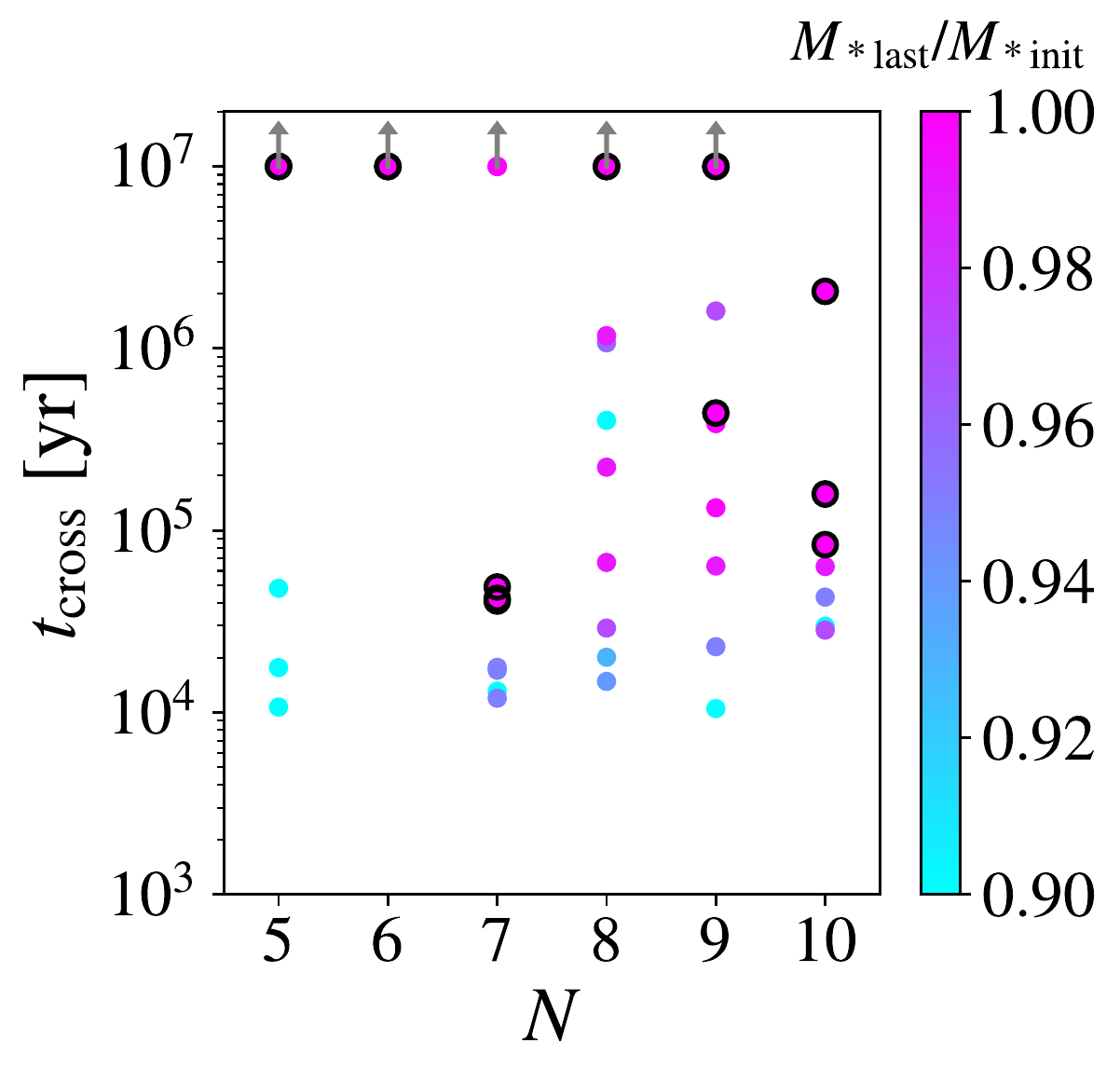}
			{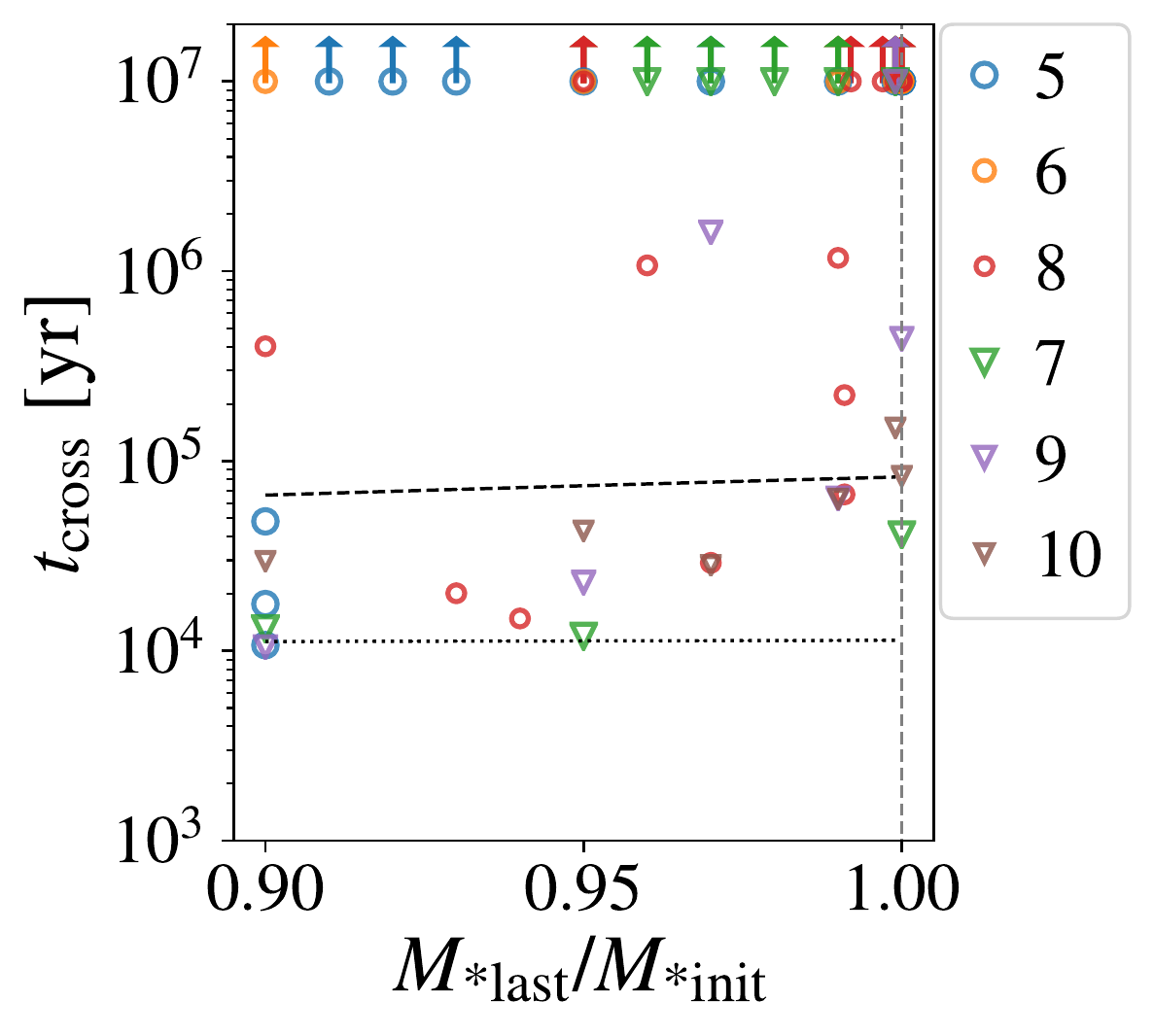}
	\caption{
		The same as Figure \ref{Fig:Mml_t_cross_4_43_Ms} but for the $\mu$5$p$4 model.
		}
	\label{Fig:Mml_t_cross_5_54_Ms}
\end{figure*}

We performed simulations in models $\mu$5$p$5 and $\mu$5$p$4, including the stellar mass evolution.
We showed the orbital crossing times in these models as functions of $N$ and $M_{\rm *last}/M_{\rm *init}$ in Figures \ref{Fig:Mml_t_cross_5_65_Ms} and \ref{Fig:Mml_t_cross_5_54_Ms}.
These figures have similar features to Figure \ref{Fig:Mml_t_cross_4_43_Ms}. 
When $N\sim N_{\rm crit}$ planets are in resonant chains, they cause orbital instabilities when stars lose a few percentages of their masses.

\bibliography{bibtex_ym}{}
\bibliographystyle{aasjournal}

\end{document}